\documentclass[twocolumn,showpacs,amsmath,amssymb]{revtex4-1}
\usepackage{graphicx}% Include figure files
\usepackage{dcolumn}% Align table columns on decimal point
\usepackage{bm}% bold math
\usepackage[abs]{overpic}
\usepackage{color}

\def\Vec#1{\bm{#1}}

\begin{document}

%\preprint{}

\title{Multi-band Eilenberger theory of superconductivity: Systematic low-energy projection}
%
%Non-local anisotropic potentials for multi-band superconductors: unified multi-band quasiclassical framework}
% with the first-principles calculations}
%%Anisotropy of Thermal Conductivity and Massive Dirac equations with a Superconductivity: 
%%an Exclusive Probe in Topological Superconductor Cu$_x$Bi$_2$Se$_3$}

\author{Yuki Nagai}
\affiliation{CCSE, Japan  Atomic Energy Agency, 178-4-4, Wakashiba, Kashiwa, Chiba, 277-0871, Japan}
\author{Hiroki Nakamura}
\affiliation{CCSE, Japan  Atomic Energy Agency, 178-4-4, Wakashiba, Kashiwa, Chiba, 277-0871, Japan}
%\author{Masahiko Machida}
%\affiliation{CCSE, Japan  Atomic Energy Agency, 5-1-5 Kashiwanoha, Kashiwa, Chiba, 277-8587, Japan}

\date{\today}% It is always \today, today,
             
\begin{abstract}
We propose the general multi-band quasiclassical Eilenberger theory of superconductivity to describe quasiparticle excitations in inhomogeneous systems.  
With the use of low-energy projection matrix, the $M$-band quasiclassical Eilenberger equations are systematically obtained from $N$-band Gor'kov equations.
Here $M$ is the internal degrees of freedom in the bands {\it crossing} the Fermi energy and $N$ is the degree of freedom in a model.
Our framework naturally includes inter-band off-diagonal elements of Green's functions, which have usually been neglected in previous multi-band quasiclassical frameworks.
The resultant multi-band Eilenberger and Andreev equations are similar to the single-band ones, except for multi-band effects. 
The multi-band effects can exhibit the non-locality and the anisotropy in the mapped systems. 
Our framework can be applied to an arbitrary Hamiltonian (e.g. a tight-binding Hamiltonian derived by the first-principle calculation).
As examples, we use our framework in various kinds of systems, such as 
noncentrosymmetric superconductor CePt$_{3}$Si, three-orbital model for Sr$_{2}$RuO$_{4}$, heavy fermion CeCoIn$_{5}$/YbCoIn$_{5}$ superlattice, 
a topological superconductor with the strong spin-orbit coupling Cu$_{x}$Bi$_{2}$Se$_{3}$, 
and a surface system on a topological insulator. 
\end{abstract}

\pacs{
%74.20.Rp, %Pairing symmetries (other than s-wave)
%74.25.Op, %Mixed states, critical fields, and surface sheaths
%74.81.-g	%Inhomogeneous superconductors and superconducting systems, including electronic inhomogeneities
%74.25.Bt  %Thermodynamic properties
}
\maketitle
%%%%%%%%%
\section{Introduction}
Multi-band superconductors such as MgB$_2$ and the iron pnictides have been attracted much attention because of its 
high critical temperature. 
Although MgB$_2$ is a phonon-mediated superconductor, it has a large critical temperature $T_c \sim 40$K, originating from the multi-band effects. 
Multi-band effects are recognized as one of the ways to increase the critical temperature. 
The discovery of the iron-pnictides had a striking impact on many researchers in condensed matter physics. 
Many kinds of phenomena with multi-band effects have been proposed and confirmed in iron-based superconductors\cite{Kobayashi,Kontani,Kuroki}. 
Furthermore, recently found superconductors characterized by topological invariants, {\it i.e.} topological superconductors, are also multi-band superconductors, 
since the internal degrees of freedom ({\it e.g.} spins, orbitals or particle-hole spaces) in a multi-band system  induce topological twists in wave functions\cite{Hasan;Kane:2010,Ando:2013,Schnyder;Ludwig:2008,Fu;Berg:2009}.  

The huge computational cost originating from the multiple degrees of freedom prevents theorists from understanding the 
physical properties in multi-band superconductors. 
For example, in the iron-based superconductor LaFeAsO, the five-orbital two-dimensional tight-binding model has been used as the effective model\cite{Kuroki}. 
There are also ten-orbital three-dimensional tight-binding models as effective models to analyze experiments in another iron-pnictides.\cite{NagaiPro}
In addition, when dealing with vortices and surfaces in multi-band systems, the computational cost becomes huger, since the momentum is not a good quantum number in inhomogeneous systems. 
For example, in the topological superconductors, it is important to study the quasiparticle excitations so-called the Majorana fermions around surfaces and vortices, in terms of the bulk-edge correspondence\cite{Hasan;Kane:2010}. 
The Ginzburg-Landau framework, which is usually used to examine the distribution of the order parameter in the inhomogeneous superconductors, 
is not suitable for dealing with the quasiparticle excitations. 
Even if we use the mean-field framework such as the Bogoliubov-de Gennes framework, the simulations of the nano-size multi-band superconductors needs enormous computational costs. 

\begin{figure}[t]
\begin{center}
\resizebox{ 0.7\columnwidth}{!}{\includegraphics{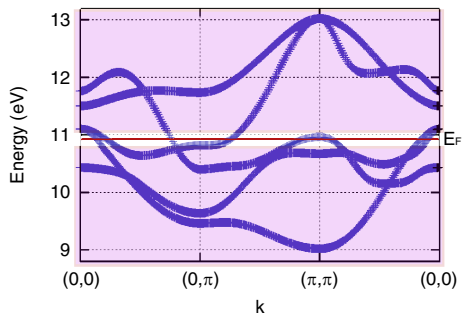}}
\end{center}
\caption{\label{fig:Feband}(Color online) Schematic figure of the multi-band Eilenberger theory. The band dispersions are calculated in the five-orbital effective model for LaOFeAs\cite{Kuroki,NagaiNJP}. The bands in the shaded regions are neglected in the multi-band Eilenberger theory. 
}
\end{figure}

We point out that the effective models ({\it e.g.}, derived by the first principles calculations) might have too many bands to describe the low energy physics of superconductivity.  
We should note that the number of the bands crossing the Fermi level in normal states is {\it less than} four even in models for iron-pnictides. 
The low energy physics in superconducting state are characterized by the quasiparticles on the bands crossing the Fermi level in normal states, 
since a characteristic energy scale of the superconducting gap ($\sim$meV) is much smaller than that of the bands  far from the Fermi level ($\sim$ eV). 
In the single-band weak-coupling Bardeen-Cooper-Schrieffer (BCS) framework, the theory using information only at the Fermi surface, called the quasiclassical Eilenberger theory, has many successes\cite{KopninText}. 
In multi-band superconductors, 
eliminating the high-energy bands not crossing the Fermi level can reduce the number of the bands in a low energy effective theory as shown in Fig.\ref{fig:Feband}, since 
the high-energy bands can not affect the physical quantities in superconducting state. 

The quasiclassical Eilenberger theory is successful in the BCS model of superconductivity. 
The theoretical framework is based on the fact that the coherence length $\xi$ is sufficiently greater than the Fermi wavelength $1/k_{\rm F}$. 
Various kinds of analytical and numerical techniques on the quasiclassical Eilenberger theory have been developed and 
successfully applied to the studies of a large number of conventional and unconventional superconductors\cite{KopninText,Volovik,NagaiJPSJ:2006,Eilenberger,NagaiMeso,Miranovic,Melnikov,NagaiPRL,NagaiCe,Graser,Iniotakis}.
The Eilenberger theory was applied into the two-band superconductor MgB$_{2}$. 
In the conventional models for MgB$_{2}$, one neglects the off-diagonal inter-band elements in Green's function.
In this case, two decoupled Eilenberger equations can describe the quasiparticle excitations, and the multi-band effects are included only through solving the gap equations\cite{Mugikura,KTanaka,Gumann,Kogan}. 

%we can choose the basis where the normal and superconducting Hamiltonian matrices become diagonal in the effective model for MgB$_{2}$.  

There are two kinds of bases to consider the multi-band systems. 
First one is the basis which is orthogonal in the momentum space (so-called ``band''-basis). 
Other is the basis which is orthogonal in the real space, such as the $d$-orbitals in models for iron-based superconductors and spins in a model with the spin-orbit coupling term. 
We call these real-space orthogonal basis ``orbital'' basis in this paper. 
In the quasiclassical Eilenberger theory, the quasiclassical Green's functions depend on the momentum of the relative motion in momentum space and the center-of-mass coordinate in real space.
In the previous multi-band quasiclassical theories\cite{Prozorov,Silaev}, 
the decoupled Eilenberger equations are used by neglecting the off-diagonal elements of the Hamiltonian in both band basis in momentums space and orbital basis in real space. 
In the case of the two-band model for MgB$_{2}$, this assumption is valid to describe the quasiparticle excitations.
%
%The decoupled Eilenberger equations can be used when the Hamiltonian matrix is simultaneous diagonal with both band basis in momentum space and orbital basis in real space.
%Thus, in the case of the two-band model for MgB$_{2}$, one can use two decoupled Eilenberger equations to describe the quasiparticle excitations in this model. 
There is, however, no Eilenberger theory which includes the off-diagonal inter-band elements, except for a perturbative approach\cite{Moor}.
The inter-band elements of Green's function are important in the complicated multi-band systems, such as the iron-pnictides and the topological superconductors.
%
%There is, however, no appropriate Eilenberger theory in the complicated multi-band systems, such as the iron-pnictides and the topological superconductors, since the Hamiltonian is not simultaneous diagonal with the band and orbital basis. 
For example, 
the iron-based superconductors have many entangled Fe-$3d$ orbitals at the Fermi level. 
The Hamiltonian proposed in the iron-based superconductors is diagonalized in momentum space by the momentum dependent unitary matrix. 
The ratio of which orbital is dominant originates from this unitary matrix and depends on the momentum. 
This ``orbital'' character, how the orbitals are entangled, at the each Fermi wave number is important to understand the physics in iron-based superconductors\cite{Kemper}. 
The off-diagonal inter-band elements of Green's functions are induced by those of the unitary matrix. 
These off-diagonal elements become important when a self-energy is induced by an inter-band scattering, which is important in a system with impurities or vortices. 
%
%
%Thus,  this ``orbital'' character, how the orbitals are entangled, at the each Fermi wave number is important to understand the physics in iron-based superconductors\cite{Kemper}. 
A model of topological superconductors usually have the off-diagonal elements in spin-space due to a spin-orbit coupling.
%
%In models of topological superconductors, 
%there is a 
%
%there is the basis called spin, which is orthogonal in a real space representation. 
%Thus, the spin basis can be regarded as the ``orbital'' basis. 
The spins rotate in momentum space, originating from the spin-orbit coupling so that 
the Hamiltonian is not diagonal with the use of spin basis in real space. 
The ``spin'' character, how the spins are entangled, induces topological superconductivity even in system with $s$-wave on-site pairing interaction in two dimension\cite{Sato,Nagai2D}. 
Therefore, the information about ``orbital'' characters in momentum space is important factor to describe multi-band effects. 

In this paper, we propose a quasiclassical Eilenberger framework in multi-band superconductors with a systematic low-energy projection. 
By eliminating the high energy bands far from the Fermi level, we derive the multi-band Andreev equations and quasiclassical Eilenberger equations 
in the projected space constructed from the bands crossing the Fermi level. 
We show that the resultant multi-band Eilenberger equations are similar to the single-band ones, except for some corrections to describe multi-band effects. 
The quasiclassical framework uses the fact that the coherence length $\xi$ is usually longer than the Fermi wave length $1/k_{\rm F}$ in a lot of superconductors. 
The orbital characters on the Fermi surfaces in normal states are naturally included in our theory.

This paper is organized as follows. 
In Sec.~\ref{sec:model}, we introduce the model of the multi-band superconductors. 
The mean-field multi-band Bogoliubov-de Gennes (BdG) Hamiltonian is proposed. 
We introduce the multi-band BdG equations and the gap equations, which is the starting point of the quasiclassical theory. 
In Sec.~\ref{sec:dec}, we discuss the decoupled Eilenberger theory used in past studies. 
We show that orbital characters can not be included in this theory. 
In Sec.~\ref{sec:wave}, we derive the multi-band quasiclassical Andreev equations, starting with the multi-band BdG Hamiltonian.
The Andreev equations describe the wave functions in the quasiclassical approach. 
In Sec.~\ref{sec:dec}, the multi-band quasiclassical Eilenberger equations are derived. 
The Eilenberger equations are the equations of motion of the quasiclassical Green's function. 
We discuss the difference between the previous theory and our theory. 
In Sec.~\ref{sec:multi}, we discuss the physical meanings of our multi-band Eilenberger theory. 
In Sec.~\ref{sec:single}, we apply the multi-band Eilenberger theory in the various kinds of systems as examples. 
We show that the previous theoretical results are reproduced by our theory and the corrections originating from orbital characters are important in multi-band systems.
In Sec.~\ref{sec:sum}, the summary is given. 

\section{Model}
\label{sec:model}
\subsection{Multi-band BdG equations}
Let us start with the mean-field BdG Hamiltonian in the $2 N \times 2 N$ Nambu-Gor'kov space, 
\begin{align}
{\cal H} &= \frac{1}{2}  \int d \Vec{r}_1 d \Vec{r}_2 \Vec{\Psi}(\Vec{r}_1)^{\dagger}
\check{H}(\Vec{r}_1,\Vec{r}_2)
\Vec{\Psi}(\Vec{r}_2). \label{eq:1}
\end{align}
Here, the column vector $\Vec{\Psi}(\Vec{r})$ is composed of $N$ fermionic annihilation $\psi_{\alpha}$ and creation operators $\psi_{\alpha}^{\dagger}$ at the position $\Vec{r}$ ($\alpha = 1,\cdots,N$), $\Vec{\Psi}(\Vec{r}) = (\{ \psi_{\alpha}(\Vec{r}) \}, \{ \psi_{\alpha}^{\dagger}(\Vec{r}) \})^{\rm T}$, where $\{ \psi_{\alpha}(\Vec{r}) \} = (\psi_{1}(\Vec{r}),\cdots, \psi_{N}(\Vec{r}))^{\rm T}$ and $\{ \psi_{\alpha}^{\dagger}(\Vec{r}) \} = (\psi_{1}^{\dagger}(\Vec{r}),\cdots, \psi_{N}^{\dagger}(\Vec{r}))^{\rm T}$. 
The subscript $\alpha$ in $\psi_{\alpha}(\Vec{r})$ or $\psi_{\alpha}^{\dagger}(\Vec{r})$ indicates a quantum index depending on spin or 
orbital, etc. 
These quantum indices are labeled by the orthogonal basis in real space, which we call orbital basis. 
The Bogoliubov-de Gennes Hamiltonian with a matrix form is composed of 
\begin{align}
\check{H}(\Vec{r}_1,\Vec{r}_2) \equiv
\check{H}^{\rm N}(\Vec{r}_1,- i \Vec{\nabla}_{1} )\delta(\Vec{r}_1-\Vec{r}_2)   + \check{\Delta}(\Vec{r}_1,\Vec{r}_2). 
\end{align}
Throughout the paper, {\it hat} $\hat{a}$ denotes a $N \times N$ matrix and {\it check} $\check{a}$ denotes a $2N \times 2N$ matrix. 
The $2 N \times 2 N$ normal-state Hamiltonian matrix $\check{H}(\Vec{r}_1,\Vec{r}_2)$ and superconducting order parameter matrix $\check{\Delta}(\Vec{r}_1,\Vec{r}_2)$ are respectively defined by 
\begin{align}
&\check{H}^{\rm N}(\Vec{r}_1,- i \Vec{\nabla}_{1}  ) \nonumber \\
&\equiv 
\left(\begin{array}{cc}
\hat{H}^{\rm N}(\Vec{r}_1,- i \Vec{\nabla}_{1}  )  & 0 \\
0 & -\hat{H}^{\ast {\rm N}}(\Vec{r}_1,- i \Vec{\nabla}_{1}  )
\end{array}\right),
\end{align}
\begin{align}
\check{\Delta}(\Vec{r}_1,\Vec{r}_2) &\equiv
\left(\begin{array}{cc}
0 & \hat{\Delta}(\Vec{r}_1,\Vec{r}_2)  \\
\hat{\Delta}^{\dagger}(\Vec{r}_2,\Vec{r}_1)  & 0
\end{array}\right).
\end{align}
The order parameter matrix is given by (so-called the gap equations) \cite{SigristUeda} 
\begin{align}
\hat{\Delta}_{\alpha \beta}(\Vec{r}_1,\Vec{r}_2) &= \sum_{\Vec{k},\Vec{q}} e^{i \Vec{k} \cdot (\Vec{r}_{1} - \Vec{r}_{2})}
e^{i \Vec{q} \cdot (\Vec{r}_{1} + \Vec{r}_{2})/2} \hat{\Delta}_{\alpha \beta} (\Vec{k},\Vec{q}), \\
\hat{\Delta}_{\alpha \beta} (\Vec{k},\Vec{q}) &= - 
\sum_{\Vec{k}', \gamma \gamma'} V_{\beta \alpha ; \gamma \gamma'}(\Vec{k},\Vec{k}') 
\langle 
\psi_{\Vec{q}/2+\Vec{k}',\gamma} \psi_{\Vec{q}/2 - \Vec{k}', \gamma'}
\rangle, \label{eq:gapeq}
\end{align}
with the multi-orbital interaction matrix $V_{\alpha \beta;\gamma \gamma'}(\Vec{k},\Vec{k}')$ and 
$\psi_{\alpha}(\Vec{r}) = \sum_{\Vec{k}} \psi_{\Vec{k},\alpha} \exp (i \Vec{k} \cdot \Vec{r})$. 

The multi-band BdG equations are then expressed as 
\begin{align}
\int d \Vec{r}_{2} \check{H}(\Vec{r}_{1},\Vec{r}_{2}) \Vec{\phi}(\Vec{r}_{2}) &= E \Vec{\phi}(\Vec{r}_{1}).  \label{eq:BdGeq}
\end{align}
With the use of the eigenvectors $\Vec{\phi}(\Vec{r}_{1})$, 
the mean-field BdG Hamiltonian (\ref{eq:1}) is diagonalized. 

\subsection{Multi-band Gor'kov equations}
The Dyson equation in Nambu-Gor'kov space (Gor'kov equation) is obtained by 
 %%%
\begin{align}
& \int d\Vec{r}' \left(i \omega_{n} \check{1}\delta(\Vec{r}_{1}- \Vec{r'}) -  \check{H}^{\rm N}(\Vec{r}_1,\Vec{r}')  \right. \nonumber \\ 
& \left. - \check{\Delta}(\Vec{r}_{1},\Vec{r}') -\check{\Sigma}(\Vec{r}_1,\Vec{r}',i \omega_n) \right)  %\nonumber \\
\check{G}(\Vec{r}',\Vec{r}_2, i \omega_n)  
= \delta(\Vec{r}_1-\Vec{r}_2) \check{1} ,
\end{align}
with 
\begin{align}
\check{H}^{\rm N}(\Vec{r}_1,\Vec{r}_2) \equiv 
\left(\check{H}^{\rm N 0}(\Vec{r}_{1}) + \check{H}^{\rm N 1}(- i  \Vec{\nabla}_{\Vec{r}_{1}}) \right) \delta(\Vec{r}_{1} - \Vec{r}_{2}) ,
\end{align}
where, $\omega_{n} = (2 n + 1) \pi T$ is the fermionic matsubara frequency, 
$\check{\Sigma}(\Vec{r}_1,\Vec{r}',i \omega_n)$ denotes the self-energy. 
Here, the $2 N \times 2 N$ Green's function is determined by 
\begin{align}
\check{G}(\Vec{r}_1,\Vec{r}_2,\tau_1-\tau_2) &\equiv - \langle {\rm T}_{\tau}
\Psi(\Vec{r}_1,\tau_1) \Psi^{\dagger}(\Vec{r}_2,\tau_2)
\rangle, \\
&= 
\left(\begin{array}{cc}
\hat{G}(\Vec{r}_1,\Vec{r}_2,\tau_1-\tau_2) & \hat{F}(\Vec{r}_1,\Vec{r}_2,\tau_1-\tau_2) \\
\hat{\bar{F}}(\Vec{r}_1,\Vec{r}_2,\tau_1-\tau_2) & \hat{\bar{G}}(\Vec{r}_1,\Vec{r}_2,\tau_1-\tau_2)
\end{array}\right), 
\end{align}
\begin{align}
G_{\alpha \beta} (\Vec{r}_1,\Vec{r}_2,\tau_1-\tau_2) &\equiv 
- \langle {\rm T}_{\tau}
\psi_{\alpha}(\Vec{r}_1,\tau_1) \psi^{\dagger}_{\beta}(\Vec{r}_2,\tau_2)\rangle,\\
F_{\alpha \beta} (\Vec{r}_1,\Vec{r}_2,\tau_1-\tau_2) &\equiv 
- \langle {\rm T}_{\tau}
\psi_{\alpha}(\Vec{r}_1,\tau_1) \psi_{\beta}(\Vec{r}_2,\tau_2)\rangle,\\
\bar{F}_{\alpha \beta} (\Vec{r}_1,\Vec{r}_2,\tau_1-\tau_2) &\equiv 
- \langle {\rm T}_{\tau}
\psi_{\alpha}^{\dagger}(\Vec{r}_1,\tau_1) \psi^{\dagger}_{\beta}(\Vec{r}_2,\tau_2)\rangle, \\
\bar{G}_{\alpha \beta} (\Vec{r}_1,\Vec{r}_2,\tau_1-\tau_2) &\equiv 
- \langle {\rm T}_{\tau}
\psi_{\alpha}^{\dagger}(\Vec{r}_1,\tau_1) \psi_{\beta}(\Vec{r}_2,\tau_2)
\rangle, 
\end{align}
with the imaginary time $\tau$. 
The local density of states is given by 
\begin{align}
N(\Vec{r},E) &=\frac{-1}{\pi} {\rm Im} \:  \left[ \lim_{\eta \rightarrow +0} {\rm Tr} \: \hat{G}(\Vec{r},\Vec{r},i \omega_{n} \rightarrow E + i \eta) \right].
\end{align}

\section{Decoupled Eilenberger equations}
\label{sec:dec}
\subsection{Model and assumptions}
Let us discuss the decoupled multi-band quasiclassical theory, which is appropriate in the conventional two-band models for MgB$_{2}$\cite{Mugikura}. 
We assume that all $N \times N$ matrices in the BdG Hamiltonian are diagonal in the ``band'' basis, expressed as 
\begin{align}
\left[ \hat{H}^{\rm N} (\Vec{r}_1,- i \Vec{\nabla}_{1}  ) \right]_{\alpha \beta} \sim  H^{\rm N}_{\alpha}(\Vec{r}_1,- i \Vec{\nabla}_{1}  ) \delta_{\alpha \beta}, \\
\left[ \hat{\Delta}(\Vec{r}_1,\Vec{r}_2) \right]_{\alpha \beta} \sim \Delta_{\alpha}(\Vec{r}_1,\Vec{r}_2) \delta_{\alpha \beta}. 
\end{align}
This assumption is equivalent to that off-diagonal inter-band elements are neglected.
%We consider the basis where the normal and superconducting Hamiltonian matrices become diagonal, since we assume that all matrices are commutative in the conventional models for MgB$_{2}$.   
In this case, the normal state Hamiltonian in momentum space is expressed as 
\begin{align}
 \hat{H}^{\rm N} (\Vec{k}) &= \left(\begin{array}{ccc}
 \lambda_{1}(\Vec{k})  & 0 & 0 \\
 0 & \ddots & 0 \\
 0 & 0 & \lambda_{N}(\Vec{k}) 
 \end{array}\right), 
\end{align}
with the eigenvalues $\lambda_{i}(\Vec{k})$. 
In real space, the Fourier transformation on each band makes the band-diagonal Hamiltonian. 
The mean-field BdG Hamiltonian in Eq.~(\ref{eq:1}) is rewritten as 
\begin{align}
{\cal H} &=  \frac{1}{2} \sum_{\alpha} \int d \Vec{r}_1 d \Vec{r}_2 \Vec{\Psi}_{\alpha}(\Vec{r}_1)^{\dagger}
\check{H}_{\alpha}(\Vec{r}_1,\Vec{r}_2)
\Vec{\Psi}_{\alpha}(\Vec{r}_2), 
\end{align}
with 
\begin{align}
&\check{H}_{\alpha}(\Vec{r}_1,\Vec{r}_2) \equiv \nonumber \\
&\left(\begin{array}{cc}
H^{\rm N}_{\alpha}(\Vec{r}_1,- i \Vec{\nabla}_{1}  )\delta(\Vec{r}_1-\Vec{r}_2)   & 
\Delta_{\alpha}(\Vec{r}_1,\Vec{r}_2)
 \\
\Delta_{\alpha}^{\ast}(\Vec{r}_2,\Vec{r}_1) & -H^{{\rm N} \ast}_{\alpha}(\Vec{r}_1,- i \Vec{\nabla}_{1})\delta(\Vec{r}_1-\Vec{r}_2)    
\end{array}\right).
\end{align}
Here, the column vector $\Vec{\Psi}_{\alpha}(\Vec{r})$ is composed of fermionic annihilation $\psi_{\alpha}$ and creation operators $\psi_{\alpha}^{\dagger}$ on the $\alpha$ band at the position $\Vec{r}$ ($\alpha = 1,\cdots,N$), $\Vec{\Psi}_{\alpha}(\Vec{r}) = (\psi_{\alpha}(\Vec{r}), \psi_{\alpha}^{\dagger}(\Vec{r}))^{\rm T}$. 
There is no inter-band effect in the Hamiltonian, since the BdG Hamiltonian $\check{H}_{\alpha}$ is determined on each band. 
The gap equations in Eq.~(\ref{eq:gapeq}) are rewritten as 
\begin{align}
 \Delta_{\alpha}(\Vec{k},\Vec{q}) &= - \sum_{\Vec{k},\gamma} V_{\alpha \gamma}(\Vec{k},\Vec{k}') 
 \langle 
\psi_{\Vec{q}/2+\Vec{k}',\gamma} \psi_{\Vec{q}/2 - \Vec{k}', \gamma}
\rangle.
\end{align}
In this approximation, the multi-band effects are included as the inter-band pairing interactions only in the pairing $V_{\alpha \gamma}$. 
Thus, the quasiclassical decoupled Eilenberger equations are easily obtained as 
\begin{align}
& i \Vec{v}_{\rm F,\alpha} \Vec{\nabla}_{\Vec{R}} \check{g}_{\Vec{R}}^{\alpha}(\Vec{k}_{\rm F},z) + 
\left[z \sigma_{z} - \check{\Delta}_{\Vec{R}}^{\alpha}(\Vec{k}_{\rm F}) \sigma_{z}
, \check{g}_{\Vec{R}}^{\alpha}(\Vec{k}_{\rm F},z) \right]_{-} = 0, \label{eq:mgb2type} \\
& \check{\Delta}_{\Vec{R}}^{\alpha}(\Vec{k}_{\rm F}) = 
- \pi T \sum_{n} \sum_{\beta} \int \frac{d S_{F}}{|\Vec{v}_{\rm F}|} V_{\alpha \beta}(\Vec{k}_{\rm F},\Vec{k}_{\rm F}') 
f_{\Vec{R}}^{\beta}(\Vec{k}_{\rm F}',i \omega_{n}), 
\end{align}
with quasiclassical Green's function on the $\alpha$ band expressed as 
\begin{align}
\check{g}_{\Vec{R}}^{\alpha}(\Vec{k}_{\rm F},z) &\equiv \oint d \xi_{\Vec{k}}^{\alpha} \sigma_{z} \check{G}^{\alpha}_{\Vec{R}}(\Vec{k},z), \\
&\equiv \left(\begin{array}{cc}
g_{\Vec{R}}^{\alpha}(\Vec{k}_{\rm F}',i \omega_{n}) & f_{\Vec{R}}^{\alpha}(\Vec{k}_{\rm F}',i \omega_{n}) \\
\bar{f}_{\Vec{R}}^{\alpha}(\Vec{k}_{\rm F}',i \omega_{n}) & \bar{g}_{\Vec{R}}^{\alpha}(\Vec{k}_{\rm F}',i \omega_{n})
\end{array}\right).
\end{align}
Here, we neglect the self energies and the vector potentials for simplicity, $z$ denotes a complex energy, $\sigma_{z}$ is the Pauli matrix in $2 \times 2$ Nambu-Gor'kov space, $[\check{A},\check{B}]_{-} = \check{A} \check{B} - \check{B} \check{A}$ is used, and 
$G^{\alpha}_{\Vec{R}}(\Vec{k},z)$ is the Green's function of the $\alpha$ band. 
It should be noted that the self-energy must be diagonalized in this basis and $\check{A}$ in this section is $2 \times 2$ matrix in Nambu space.  
The above equations were used in the simple multi-band superconductors such as MgB$_{2}$. 
For MgB$_{2}$, there are two bands (so-called $\sigma$- and $\pi$-bands) and the gap equations connect information on each band. 

\subsection{Appropriate region of the decoupled Eilenberger equations}
Now, we discuss the appropriate region of the decoupled Eilenberger equations in the previous section. 
We introduce the Hamiltonians $\hat{H}^{\rm orbital}(\Vec{k})$ and $\hat{H}^{\rm band}(\Vec{k})$ in normal states with the orbital basis and the band basis in momentum space, respectively.  
Generally, the Hamiltonian $\hat{H}^{\rm orbital}(\Vec{k})$  has the off-diagonal elements, since the orbital basis is a basis which is diagonal in real space.  
With the use of the momentum dependent unitary matrix $\hat{U}(\Vec{k})$, 
one can obtain the diagonal Hamiltonian expressed as 
\begin{align}
\hat{U}^{\dagger}(\Vec{k}) \hat{H}^{\rm orbital}(\Vec{k}) \hat{U}(\Vec{k}) 
&= \left(\begin{array}{ccc}
 \lambda_{1}(\Vec{k})  & 0 & 0 \\
 0 & \ddots & 0 \\
 0 & 0 & \lambda_{N}(\Vec{k}) 
 \end{array}\right), \\
&\equiv \hat{H}^{\rm band}(\Vec{k}).
\end{align}
The ``band'' basis is determined by the unitary transformation of the orbital basis in momentum space. 
If the unitary matrix does not depend on momentum, one can choose the basis which simultaneously diagonalizes the Hamiltonian in both real and momentum spaces. 
Generally, the decoupled Eilenberger equations are derived by neglecting the off-diagonal elements.

We show three examples that the decoupled Eilenberger theory is not appropriate as follows. 
The first example is an impurity problem in the multi-band superconductors. 
In the previous study\cite{Belova}, they assumed that the impurity-induced self energy was described as a momentum-independent band-diagonal matrix, which lead to the decoupled Eilenberger equations.
We point out that this assumption induces non-local impurities in real space, as follows.
In the band basis with this assumption, the Gor'kov equations become 
\begin{align}
\check{G}_{\Vec{k}}^{\rm band}(z)&= \check{G}_{\Vec{k}}^{0 {\rm band}}(z) + \check{G}_{\Vec{k}}^{0 {\rm band}}(z) \check{\Sigma}^{\rm band}(z) \check{G}_{\Vec{k}}^{\rm band}(z).
\end{align}
Here, $\check{A}^{\rm band}$ denotes the matrix defined by the basis which diagonalizes the normal-state Hamiltonian $\check{H}_{\Vec{k}}^{\rm orbital}$. 
To describe impurities in real space, one has to use the orbital basis in real space.
%When considering what kinds of impurities are in real space, one has to use the orbital basis determined in real space.
With use of the unitary transformation from the band basis to the orbital basis, 
the impurity-induced self-energy in the orbital basis $\check{\Sigma}^{{\rm orbital}}$ should have a momentum dependence expressed as 
\begin{align}
\check{\Sigma}^{{\rm orbital}}(\Vec{k},z) &= \check{U}(\Vec{k})  \check{\Sigma}^{\rm band}(z)\check{U}^{\dagger}(\Vec{k}),
\end{align}
with 
\begin{align}
\check{U}(\Vec{p}) &= \left(\begin{array}{cc}\hat{U}(\Vec{p}) & 0 \\0 & \hat{U}^{\ast}(-\Vec{p})\end{array}\right). 
\end{align}
If the self-energy in the band basis is obtained by the $T$-matrix approximation for randomly distributed point impurities given as 
\begin{align}
\check{\Sigma}^{{\rm band}}(z) &=n_{\rm imp}\check{V} + n_{\rm imp}   \sum_{\Vec{p}} \check{V}  \check{G}_{\Vec{p}}^{0 \: {\rm orbital}}(z)  \check{V},
\label{eq:sigma2}
\end{align}
the self-energy in the orbital basis becomes 
\begin{align}
&\check{\Sigma}^{\rm orbital}(\Vec{k}, z) =n_{\rm imp} \check{V}^{\rm orbital}(\Vec{k},\Vec{k}) \nonumber \\
&+   n_{\rm imp}  \sum_{\Vec{p}} 
 \check{V}^{\rm orbital}(\Vec{k},\Vec{p})
 \check{G}_{\Vec{p}}^{0 {\rm orbital}}(z)   \check{V}^{\rm orbital}(\Vec{p},\Vec{k}),
\end{align}
with the effective ``non-local'' impurity potential defined as 
\begin{align}
 \hat{V}^{\rm orbital}(\Vec{k},\Vec{p}) &\equiv \check{U}(\Vec{k}) \check{V} \check{U}^{\dagger}(\Vec{p}).
\end{align}
Therefore, the decoupled Eilenberger equations does not describe the local impurity potentials.

The second example is the proximity-induced impurity-robust $p$-wave effective order parameter on the surface of a topological insulator, as discussed later in Sec.~\ref{sec:rp}. 
With the use of the band basis, an effective chiral $p$-wave order parameter can be derived by the previous quasiclassical framework. 
This previous framework, however, can not describe the robustness against non-magnetic impurities, which was proposed by directly solving the BdG equations\cite{Ito:2011ct}.
The impurity robust $p$-wave superconductor is naturally introduced in our framework.

The third example is the appearance condition of the Andreev bound states at a surface. 
In a single band model, the Andreev bound states occur when the sign of the gap function changes through the scattering process\cite{Kashiwaya:2000ic}. 
In a multi band model, an ambiguity of the "sign" of the order parameter makes the above statement unclear. 
This ambiguity can not be resolved by the previous quasiclassical framework. 
In our multi-band quasiclassical Eilenberger approach, we can overcome this difficulty by deriving the most appropriate effective order parameter, 
which obeys the statement of the Andreev bound states as discussed later in Sec.~\ref{sec:arbit}.

We can use the decoupled Eilenberger equations if we assume the momentum-independent unitary matrix $(\check{U}(\Vec{k}) = \check{U})$. 
In this assumption, however, we can not treat the ``orbital'' character. 
Therefore, we propose the general multi-band Eilenberger theory.

\section{Quasiclassical treatment I: wave-function approach}
\label{sec:wave}
In this section, we derive the quasiclassical equations on the basis of the BdG equations. 
The quasiclassical theory is founded on an assumption that the coherence length $\xi$ is much longer than the 
Fermi wave length $1/k_{\rm F}$ (i.e. $\xi k_{\rm F} \ll 1$)\cite{Volovik}. 
This assumption is valid, if the order parameter amplitude is much smaller than the Fermi energy, and 
this condition is fully fulfilled in BCS weak-coupling superconductivity. 
In this theory, the wave function is expressed by a product of the fast oscillating one characterized by the Fermi momentum 
$p_{\rm F}$ and the slowly varying one by the coherence length. 
We proposed the quasiclassical theory for the multi-orbital topological superconductor\cite{NagaiTopo}.  
The generalization of this theory is proposed in this section. 

\subsection{Assumptions}
We assume that the eigen vector $\Vec{\phi}(\Vec{r})$ in Eq. (\ref{eq:BdGeq}) is 
expressed by a product of the fast oscillating one characterized by the Fermi momentum and the slowly varying one by the coherence length expressed as 
\begin{align}
\Vec{\phi}(\Vec{r}) &= \sum_{\Vec{k}_{\rm F}} e^{i \Vec{k}_{\rm F} \Vec{r}} 
\Vec{\phi}_{\Vec{k}_{\rm F}}'(\Vec{r}), \label{eq:wave}
\end{align}
where 
\begin{align}
\Vec{\phi}_{\Vec{k}_{\rm F}}'(\Vec{r}) &\equiv 
\sum_{l=1}^{M} 
\left(\begin{array}{c}
\Vec{u}^{\rm N}_{l}(\Vec{k}_{\rm F}) f_{l}^{\Vec{k}_{\rm F}}(\Vec{r}) \\
\Vec{v}^{\rm N}_{l}(\Vec{k}_{\rm F}) g_{l}^{\Vec{k}_{\rm F}}(\Vec{r}) 
\end{array}\right).
\end{align}
Here, $f_{l}^{\Vec{k}_{\rm F}}(\Vec{r})$ and $g_{l}^{\Vec{k}_{\rm F}}(\Vec{r})$ correspond to slowly varying components, 
$\Vec{u}^{\rm N}_{l}(\Vec{k}_{\rm F})$, $\Vec{v}^{\rm N}_{l}(\Vec{k}_{\rm F})$ are 
the fast oscillating functions adopted as normal-state uniform eigenvectors satisfying the eigen-equations, 
\begin{align}
\check{H}^{\rm N1}(\Vec{k}) 
\left(\begin{array}{c}
\Vec{u}^{\rm N}_{l}(\Vec{k})\\
\Vec{v}^{\rm N}_{l}(\Vec{k}) 
\end{array}\right) &= \epsilon_{l}(\Vec{k})
\left(\begin{array}{c}
\Vec{u}^{\rm N}_{l}(\Vec{k}) \\
\Vec{v}^{\rm N}_{l}(\Vec{k})
\end{array}\right),  
\end{align}
where 
\begin{align}
\check{H}^{\rm N 1}(\Vec{k}) &= \left(\begin{array}{cc}
\hat{H}^{\rm N 1}(\Vec{k}) & 0 \\
0 & \hat{H}^{\rm N 1}(-\Vec{k})^{\ast} 
\end{array}\right).
\end{align}
The Fermi surfaces in normal states are expressed by the set of the zero-energy eigenvalues of $\hat{H}^{\rm N 1}(\Vec{k})$. 
The $M$-eigenvalues $\epsilon_{l}(\Vec{k})$ cross the Fermi level (i.~e.~ $\epsilon_{l} (\Vec{k}_{\rm F}) = 0$). 
We assume that the eigenvalues near the Fermi level  are same ($\epsilon_{1}(\Vec{k}), \cdots, \epsilon_{M}(\Vec{k}) = \xi(\Vec{k})$) expressed as 
\begin{align}
\check{H}^{\rm N1}(\Vec{k}) 
\left(\begin{array}{c}
\Vec{u}^{\rm N}_{l}(\Vec{k})\\
\Vec{v}^{\rm N}_{l}(\Vec{k}) 
\end{array}\right) &= \xi(\Vec{k})
\left(\begin{array}{c}
\Vec{u}^{\rm N}_{l}(\Vec{k}) \\
\Vec{v}^{\rm N}_{l}(\Vec{k})
\end{array}\right).  \label{eq:eigen}
\end{align}
\begin{figure}[t]
\begin{center}
\resizebox{1 \columnwidth}{!}{\includegraphics{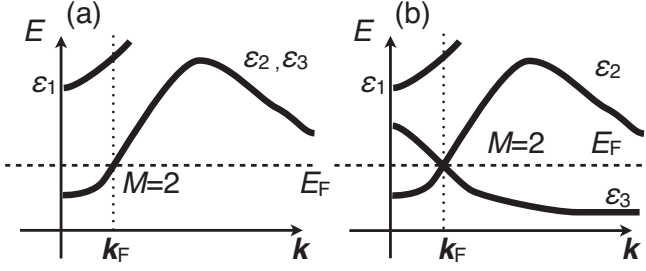}}
\end{center}
\caption{\label{fig:cross}The schematic figures of the electron bands in the multi-orbital system characterized by $M = 2$ at the Fermi wave momentum $\Vec{k}_{\rm F}$. 
}
\end{figure}
This assumption is appropriate when $M$ denotes the internal degrees of freedom at the Fermi level as shown in Fig.~\ref{fig:cross}(a). 
%
%$M$ is the internal degrees of freedom at the band crossing the Fermi energy 
We note that there is an exception as shown in Fig.~\ref{fig:cross}(b). 
For example, this exception occurs when the Fermi level is located at the center of the Dirac cone. 
However, the assumption is usually appropriate in many realistic materials.  
We should note that there is the relation between $\Vec{u}_{l}^{\rm N}(\Vec{k})$  and $\Vec{v}_{l}^{\rm N}(\Vec{k})$ at the Fermi energy expressed as 
\begin{align}
\Vec{v}_{l}^{\rm N}(\Vec{k}_{\rm F}) &= \Vec{u}_{l}^{\rm N \ast}(-\Vec{k}_{\rm F}).
\end{align}
\subsection{Andreev-type equations}
\begin{widetext}
By substituting Eq.(\ref{eq:wave}) into the BdG equations (\ref{eq:BdGeq}), 
we obtain the Andreev-type quasiclassical equations. 
Eventually, we have the $2 M \times 2 M$ quasiclassical BdG equations represented as (in detail, see Appendix \ref{app:and}),
%%%%%
\begin{align}
\left(\begin{array}{cc}
- i \Vec{v}_{\rm F} \cdot \Vec{\nabla} + V_{0}(\Vec{r},\Vec{k}_{\rm F}) & \Delta_{\rm eff}(\Vec{r},\Vec{k}_{\rm F}) \\
 \Delta_{\rm eff}^{\dagger}(\Vec{r},\Vec{k}_{\rm F}) &  i \Vec{v}_{\rm F} \cdot \Vec{\nabla} - V_{0}^{\ast}(\Vec{r},-\Vec{k}_{\rm F}) 
\end{array}\right)
\left(\begin{array}{c}
\Vec{f}^{\Vec{k}_{\rm F}}(\Vec{r}) \\
\Vec{g}^{\Vec{k}_{\rm F}}(\Vec{r})
\end{array}\right)
&=
E
\left(\begin{array}{c}
\Vec{f}^{\Vec{k}_{\rm F}}(\Vec{r}) \\
\Vec{g}^{\Vec{k}_{\rm F}}(\Vec{r})
\end{array}\right),\label{eq:and}
\end{align}
\end{widetext}
with introducing $M \times M$ matrices $V_{0}$ and $\Delta_{\rm eff}$ defined by 
%%%%%%%
\begin{align}
V_{0}(\Vec{r},\Vec{k}_{\rm F}) &\equiv \tilde{U}_{\Vec{k}_{\rm F}}^{M \dagger}  \hat{H}^{\rm N 0}(\Vec{r}) \tilde{U}_{\Vec{k}_{\rm F}}^{M}, \\
 \Delta_{\rm eff}(\Vec{r},\Vec{k}_{\rm F}) &\equiv  \tilde{U}_{\Vec{k}_{\rm F}}^{M \dagger} \hat{\Delta}(\Vec{r},\Vec{k}_{\rm F})  \tilde{U}_{-\Vec{k}_{\rm F}}^{M \ast}, \label{eq:gapeff}
\end{align}
%%%%%%
with the $N \times M$ matrix given by 
\begin{align}
\tilde{U}_{\Vec{k}_{\rm F}}^{M} = (\Vec{u}_{1}^{\rm N}(\Vec{k}_{\rm F}), \cdots, \Vec{u}_{M}^{\rm N}(\Vec{k}_{\rm F})).  \label{eq:ukm}
\end{align} 
Here, the $M$-component vectors $\Vec{f}^{\Vec{k}_{\rm F}}(\Vec{r})$ and $\Vec{g}^{\Vec{k}_{\rm F}}(\Vec{r})$ denote $\Vec{f}^{\Vec{k}_{\rm F}}(\Vec{r})^{\rm T} = ( f_{1}^{\Vec{k}_{\rm F}}(\Vec{r}), \cdots,  f_{M}^{\Vec{k}_{\rm F}}(\Vec{r}) )$ and $\Vec{g}^{\Vec{k}_{\rm F}}(\Vec{r})^{\rm T} = (g_{1}^{\Vec{k}_{\rm F}}(\Vec{r}), \cdots,  g_{M}^{\Vec{k}_{\rm F}}(\Vec{r}) )$, respectively.  
The $N \times M$ matrix has the relation expressed as 
\begin{align}
\tilde{U}_{\Vec{k}}^{M \dagger} \tilde{U}_{\Vec{k}}^{M} &= 1_{M \times M}.
\end{align}
Note that $\tilde{U}_{\Vec{k}}^{M} \tilde{U}_{\Vec{k}}^{M \dagger} \neq 1_{N \times N}$ if $M$ is not equal to $N$. 
With the use of the $N \times M$ matrix $ \tilde{U}_{\Vec{k}}^{M}$, the eigenvector $\Vec{\phi}(\Vec{r})$ in the BdG equations (\ref{eq:BdGeq}) is approximated as 
\begin{align}
\Vec{\phi}(\Vec{r}) \sim \sum_{\Vec{k} = \Vec{k}_{\rm F}} e^{i \Vec{k}_{\rm F} \Vec{r}} 
\left(\begin{array}{c}  \tilde{U}_{\Vec{k}}^{M} \Vec{f}^{\Vec{k}_{\rm F}}(\Vec{r})\\
 \tilde{U}_{-\Vec{k}}^{M \ast} \Vec{g}^{\Vec{k}_{\rm F}}(\Vec{r})
\end{array}\right).
\end{align}

The resultant quasiclassical BdG equations (\ref{eq:and}) are equivalent to the linearized BdG (Andreev) equations if we consider the 
single band system ($N = 1$)\cite{Volovik}.  
Thus, we successfully reduce the matrix dimension from the number of the bands $N$ to the number of the degenerated Fermi levels $M$ in this quasiclassical treatment. 

%%%%%
\subsection{Boundary condition at a specular surface}
Let us discuss the boundary condition at a specular surface. 
For simplicity, we consider that the material is filled in the region $z > 0$. 
By assuming the translational symmetry along the $x$ and $y$ axes which conserves the momentum $\Vec{k}_{{\rm F} \parallel} = (k_{{\rm F}x},k_{{\rm F}y})$, the boundary condition is given by 
%%%%%
\begin{align}
\Vec{\phi}(\Vec{k}_{{\rm F} \parallel},z=0) &= 0. \label{eq:bound}
\end{align}
%%%%
First, we find the solutions which satisfy the above boundary condition in normal states at the Fermi energy, expressed as 
\begin{align}
\Vec{u}^{\rm N} (\Vec{k}_{{\rm F} \parallel},z=0)&= 0. 
\end{align}
By solving the eigenvalue equations with the normal-state $N \times N$ Hamiltonian $\hat{H}^{\rm N 1}(\Vec{k})$: 
\begin{align}
\hat{H}^{\rm N 1}(\Vec{k}_{{\rm F} \parallel},k_{z}^{i}) \Vec{u}_{l,(\Vec{k}_{{\rm F} \parallel},k_{z}^{i})}^{{\rm N}} &= 0, \label{eq:boundnh}
\end{align}
the boundary condition becomes 
\begin{align}
\sum_{i}^{K} \sum_{l}^{M} c_{i}^{l}(\Vec{k}_{{\rm F} \parallel}) \Vec{u}_{l,(\Vec{k}_{{\rm F} \parallel},k_{z}^{i})}^{{\rm N}} &= 0. \label{eq:nbound}
\end{align}
Note that $k_{z}^{i}$ is a complex number and satisfies ${\rm Im} \: k_{z}^{i} \geq 0 $. 
Here, $K$ denotes the number of $k_{z}^{i}$ with the same conserved momentum ($\Vec{k}_{{\rm F} \parallel}$). 
For example, 
in the case of $K = 2$ and $M = 1$, the boundary condition is satisfied only when the vector $\Vec{u}_{2,(\Vec{k}_{{\rm F} \parallel},k_{z})}^{{\rm N}}$ 
is parallel to the vector $\Vec{u}_{1,(\Vec{k}_{{\rm F} \parallel},k_{z})}^{{\rm N}}$ expressed as 
\begin{align}
\Vec{u}_{2,(\Vec{k}_{{\rm F} \parallel},k_{z})}^{{\rm N}} &= e^{i \Phi_{12}} \Vec{u}_{1,(\Vec{k}_{{\rm F} \parallel},k_{z})}^{{\rm N}}. \label{eq:u12}
\end{align}
Here $\Phi_{12}$ is the overall phase difference between these two vectors. 
Thus, we obtain 
\begin{align}
c_{2} &= - e^{- i \Phi_{12}} c_{1} \label{eq:k2m1}.
\end{align}

In a superconducting state, we use the quasiclassically-approximated wavefunctions. 
In the quasiclassical treatment, the eigenvector $\Vec{\phi}(\Vec{k}_{{\rm F} \parallel}, \Vec{r})$ in the BdG equations is approximated as 
\begin{align}
\Vec{\phi}(\Vec{k}_{{\rm F} \parallel},\Vec{r}) \sim  \sum_{i}^{K} e^{i k_{z}^{i} z}
\left(\begin{array}{c}
\tilde{U}_{(\Vec{k}_{{\rm F} \parallel},k_{z}^{i})}^{M} \Vec{f}^{k_{z}^{i}}(\Vec{k}_{{\rm F} \parallel},z)
 \\
\tilde{U}_{(-\Vec{k}_{{\rm F} \parallel},-k_{z}^{i})}^{M \ast} \Vec{g}^{k_{z}^{i}}(\Vec{k}_{{\rm F} \parallel},z)
\end{array}\right),
\end{align}
with the boundary conditions 
\begin{align}
\sum_{i}^{K} \sum_{l}^{M} c_{i}^{l}(\Vec{k}_{{\rm F} \parallel}) f_{l}^{k_{{\rm F}z}^{i}}(\Vec{k}_{{\rm F} \parallel},z=0) &= 0, \label{eq:boundf} \\
\sum_{i}^{K} \sum_{l}^{M} c_{i}^{l \ast}(-\Vec{k}_{{\rm F} \parallel}) g_{l}^{k_{{\rm F}z}^{i}}(\Vec{k}_{{\rm F} \parallel},z=0) &= 0.\label{eq:boundg} 
\end{align}
Here, we define the $N \times M$ matrix given by 
\begin{align}
\tilde{U}_{(\Vec{k}_{{\rm F} \parallel},k_{z}^{i})}^{M} = (\Vec{u}_{1,(\Vec{k}_{{\rm F} \parallel},k_{z}^{i})}^{{\rm N}}, \cdots, \Vec{u}_{M,(\Vec{k}_{{\rm F} \parallel},k_{z}^{i})}^{{\rm N}}).  
\end{align} 
In order to find the coefficients $c_{i}^{l}(\Vec{k}_{{\rm F} \parallel})$, we have to solve the boundary condition (\ref{eq:nbound}) in normal states. 
For example, in the case of $K = 2$, the boundary condition in Eq.~(\ref{eq:nbound}) is expressed as 
\begin{align}
\tilde{U}_{(\Vec{k}_{{\rm F} \parallel},k_{z}^{1})}^{M} \Vec{c}_{(\Vec{k}_{{\rm F} \parallel},k_{z}^{1})}^{M} + 
\tilde{U}_{(\Vec{k}_{{\rm F} \parallel},k_{z}^{2})}^{M} \Vec{c}_{(\Vec{k}_{{\rm F} \parallel},k_{z}^{2})}^{M}
 &= 0, \label{eq:k2}
 \end{align}
with $\Vec{c}_{(\Vec{k}_{{\rm F} \parallel},k_{z}^{i})}^{M T} = (c_{i}^{1}(\Vec{k}_{{\rm F} \parallel}),\cdots,c_{i}^{M}(\Vec{k}_{{\rm F} \parallel}))$. 
We show the boundary conditions of two superconducting systems with $K = 2$, which can be easily obtained, as examples. 
In the case of $N = M$, the solution of the boundary condition (\ref{eq:k2}) is 
\begin{align}
\Vec{c}_{(\Vec{k}_{{\rm F} \parallel},k_{z}^{2})}^{M} &= 
-\tilde{U}_{(\Vec{k}_{{\rm F} \parallel},k_{z}^{2})}^{M \dagger}  \tilde{U}_{(\Vec{k}_{{\rm F} \parallel},k_{z}^{1})}^{M} \Vec{c}_{(\Vec{k}_{{\rm F} \parallel},k_{z}^{1})}^{M},  \label{eq:bbk2}
\end{align}
since the relation $\tilde{U}_{(\Vec{k}_{{\rm F} \parallel},k_{z}^{2})}^{M} \tilde{U}_{(\Vec{k}_{{\rm F} \parallel},k_{z}^{2})}^{M \dagger} = 1_{N \times N}$ 
is satisfied. 
Note that Eq.~(\ref{eq:bbk2}) is not the solution in Eq.~(\ref{eq:k2}) if $N \neq M$. 
By substituting Eq.~(\ref{eq:bbk2}) into Eqs.~(\ref{eq:boundf}) and (\ref{eq:boundg}), we obtain 
\begin{align}
 \Vec{f}^{k_{z}^{2}}(z=0) &= - \tilde{V}^{(\Vec{k}_{{\rm F} \parallel},k_{z}^{2})}_{(\Vec{k}_{{\rm F} \parallel},k_{z}^{1})} \Vec{f}^{k_{z}^{1}}(z=0), \label{eq:fnm} \\
 \Vec{g}^{k_{z}^{2}}(z=0)  &= -\tilde{V}^{(-\Vec{k}_{{\rm F} \parallel},-k_{z}^{2}) \ast}_{(-\Vec{k}_{{\rm F} \parallel},-k_{z}^{1})} \Vec{g}^{k_{z}^{1}}(z=0), \label{eq:gnm}
\end{align}
where the ``transfer matrix'' $\tilde{V}^{(\Vec{k}_{{\rm F} \parallel},k_{z}^{2})}_{(\Vec{k}_{{\rm F} \parallel},k_{z}^{1})}$ is determined as 
\begin{align}
\tilde{V}^{(\Vec{k}_{{\rm F} \parallel},k_{z}^{2})}_{(\Vec{k}_{{\rm F} \parallel},k_{z}^{1})}
&\equiv 
\tilde{U}_{(\Vec{k}_{{\rm F} \parallel},k_{z}^{2})}^{M \dagger} \tilde{U}_{(\Vec{k}_{{\rm F} \parallel},k_{z}^{1})}^{M}. \label{eq:transfer}
\end{align}
Next, we consider the case of $M = 1$ and $K = 2$. 
The boundary condition (\ref{eq:k2m1}) becomes 
\begin{align}
f^{k_{z}^{2}}(\Vec{k}_{{\rm F} \parallel},z=0) &= - e^{- i \Phi_{12}} f^{k_{z}^{1}}(\Vec{k}_{{\rm F} \parallel},z=0), \label{eq:fgbf} \\
g^{k_{z}^{2}}(\Vec{k}_{{\rm F} \parallel},z=0) &= - e^{i \Phi_{12}} g^{k_{z}^{1}}(\Vec{k}_{{\rm F} \parallel},z=0), \label{eq:fgb}
\end{align}
which is equivalent to the boundary condition in the single band case when $\Phi_{12} = 0$. 
We should note again that the above boundary condition can be only used when the normal-state eigenvectors with different momenta $k_{z}^{1}$ and $k_{z}^{2}$ are parallel
to each other shown in Eq.~(\ref{eq:u12}), since we have to find the correct boundary condition for $f^{k_{z}^{2}}(\Vec{k}_{{\rm F} \parallel},z=0)$ which satisfies Eq.~(\ref{eq:boundf}). 
 
The characteristic momentum $k_{z}^{i}$ is obtained by solving a normal-state eigenvalue equation (\ref{eq:boundnh}) with the boundary condition (\ref{eq:nbound}) at the Fermi energy. 
This momentum $k_{z}^{i}$ is usually a real number since wavefunctions with a Fermi wave number can usually satisfy the boundary condition. 
In this case, we solve the Andreev equations (\ref{eq:and}) with the real-number momentum $(\Vec{k}_{{\rm F} \parallel},k_{z}^{i})$. 
However, the momentum $k_{z}^{i}$ is not the Fermi wave number, when there are bound states in normal states. 
The normal-state wavefunction localized at a boundary is described by a complex momentum. 
We will show the system which needs the complex wave number to describe the bound states, as discussed in Sec.~\ref{sec:single}. 

We discuss the existence condition of solutions in Eq.~(\ref{eq:nbound}). 
With the use of the matrix representation, the equation (\ref{eq:nbound}) is rewritten as 
\begin{align}
{\cal U} \Vec{C} &= 0,
\end{align}
where the $N \times M K$ matrix ${\cal U}$ and $M K$-dimension vector $\Vec{C}$ are defined as  
\begin{align}
{\cal U} &\equiv (\tilde{U}_{(\Vec{k}_{{\rm F} \parallel},k_{z}^{1})}^{M},\cdots, \tilde{U}_{(\Vec{k}_{{\rm F} \parallel},k_{z}^{K})}^{M}  ), \\
\Vec{C} &\equiv \left(\begin{array}{c}
\Vec{c}_{(\Vec{k}_{{\rm F} \parallel},k_{z}^{1})}^{M} \\
\vdots \\
\Vec{c}_{(\Vec{k}_{{\rm F} \parallel},k_{z}^{K})}^{M}
\end{array}\right).
\end{align}
The above equations are linear homogeneous equations with $M K$ unknowns. 
These equations can have the solutions when $M K > N$.

Finally, we note that the general boundary condition for the conventional quasiclassical equations has been discussed by several groups\cite{Shelankov:2000jq,Eschrig:2009bz}. 
The incoming and outgoing wavefunctions are connected by the scattering matrix $\hat{S}$ expressed as 
\begin{align}
\Vec{\phi}(\Vec{k}_{\rm out}) &= \hat{S}_{\Vec{k}_{\rm out} \Vec{k}_{\rm in}} \Vec{\phi}(\Vec{k}_{\rm in}),
\end{align}
where $\Vec{k}_{\rm in(out)}$ is the wave number of the incoming (outgoing) quasiparticles. 
Here, $\hat{S}$ is the scattering matrix defined at the Fermi energy in normal states\cite{Eschrig:2009bz}. 
The above relation can be used to develop the general boundary condition in the multi-band superconductors. 
The development of the general boundary condition for the multi-band quasiclassical equations is a future issue.

\section{Quasiclassical treatment II: Green's function approach}
\label{sec:green}
In this section, we derive the equation of motion of the quasiclassical Green's function so-called Eilenberger equations in a multi-band system. 
The size of the matrices of the Green's function is reduced by the low-energy projection. 

 \subsection{Wigner representation}
\label{sec:wigner}
The Wigner representation is usually introduced in terms of the derivation of the quantum Boltzmann equations.  
The transport-like equations of motions of the quasiclassical Green's functions in the single band system 
are systematically derived with the use of the Wigner representations.  
We introduce the Wigner representation defined by 
\begin{align}
\check{A}(\Vec{R},\Vec{k}) &\equiv \int d\Vec{\bar{r}} e^{-i \Vec{k} \cdot \Vec{\bar{r}}} 
\check{A}(\Vec{r}_{1},\Vec{r}_{2}) \Bigl|_{\Vec{r}_{1}=\Vec{R}+\bar{\Vec{r}}/2,\Vec{r}_{2}=\Vec{R}-\bar{\Vec{r}}/2}.
\end{align}
Here, $\Vec{R} = (\Vec{r}_{1}+\Vec{r}_{2})/2$ and $\Vec{\bar{r}} = \Vec{r}_{1} - \Vec{r}_{2}$ are the center-of-mass coordinate and the relative 
coordinate, respectively.

The Gor'kov equations in the Wigner representation are expressed by 
\begin{align}
& \left(
i \omega_{n} - \check{H}^{\rm N 0}(\Vec{R}) - \check{H}^{\rm N 1}(\Vec{k}) - \check{\Delta}(\Vec{R},\Vec{k}) - \check{\Sigma}(\Vec{R},\Vec{k},i \omega_{n})
 \right) \nonumber \\
 &\star \check{G}(\Vec{R},\Vec{k},i \omega_{n}) = \check{1}. \label{eq:wig}
\end{align}
Here, we introduce the $\star$-product (Moyal product) determined by 
\begin{align}
\check{A}(\Vec{R},\Vec{k}) \star \check{B}(\Vec{R},\Vec{k}) &\equiv 
\exp \left[ \frac{i}{2 } (\Vec{\nabla}_{\Vec{k}'} \cdot \Vec{\nabla}_{\Vec{R}}-\Vec{\nabla}_{\Vec{k}} \cdot \Vec{\nabla}_{\Vec{R}'})\right] \nonumber \\
&\times \check{A}(\Vec{R},\Vec{k}) \check{B}(\Vec{R}',\Vec{k}') \Big|_{\Vec{k}' = \Vec{k}, \Vec{R}' = \Vec{R}}. 
\end{align}
We note that there is another Gor'kov equation called the ``right-hand Gor'kov'' equation. 
In terms of the Wigner representation, the right-hand equation is expressed as 
\begin{align}
& \check{G}(\Vec{R},\Vec{k},i \omega_{n}) \star \nonumber \\
&  \left(
i \omega_{n} - \check{H}^{\rm N 0}(\Vec{R}) - \check{H}^{\rm N 1}(\Vec{k}) - \check{\Delta}(\Vec{R},\Vec{k}) - \check{\Sigma}(\Vec{R},\Vec{k},i \omega_{n})
 \right) 
 = \check{1}.
\end{align}
The local density of states with the Wigner representation is expressed as 
\begin{align}
N(\Vec{r},E) &= \frac{-1}{\pi} {\rm Im} \: \left[  \lim_{\eta \rightarrow +0} \sum_{\Vec{k}} {\rm Tr} \: \hat{G}(\Vec{r},\Vec{k},i \omega_{n} \rightarrow E + i \eta) \right]. \label{eq:wigldos}
\end{align}

Let us derive the quasiclassical equations from Eq.~(\ref{eq:wig}) as follows. 
In superconductors, the characteristic length of the center-of-mass coordinates is the coherence length $\xi$, which 
is much longer than that of the relative coordinates characterized by $1/k_{\rm F}$.  
Assuming that the characteristic coherence length is long, the Moyal product in the first order of $\Vec{\nabla}_{\Vec{R}}$ is given as 
\begin{align}
& \check{A}(\Vec{R},\Vec{k}) \star \check{B}(\Vec{R},\Vec{k}) \sim
\check{A}(\Vec{R},\Vec{k}) \check{B}(\Vec{R},\Vec{k}) \nonumber \\
&+ \frac{i}{2} 
\left( 
\Vec{\nabla}_{\Vec{R}} \check{A}(\Vec{R},\Vec{k}) \cdot  \Vec{\nabla}_{\Vec{k}} \check{B}(\Vec{R},\Vec{k}) 
- \Vec{\nabla}_{\Vec{k}} \check{A}(\Vec{R},\Vec{k}) \cdot \Vec{\nabla}_{\Vec{R}} \check{B}(\Vec{R},\Vec{k}) 
\right).
\end{align}
Then, the Gor'kov equations are expressed as (see, Appendix \ref{sec:expansion})
\begin{widetext}
%%%%
\begin{align}
& \left(
i \omega_{n} - 
\check{H}^{\rm N 0}(\Vec{R})
 - \check{H}^{\rm N 1}(\Vec{k})
  - \check{\Delta}(\Vec{R},\Vec{k}) - \check{\Sigma}(\Vec{R},\Vec{k},i \omega_{n})
 \right) 
 \check{G}(\Vec{R},\Vec{k},i \omega_{n}) \nonumber \\
& - \frac{i}{2 } 
 \Vec{\nabla}_{\Vec{R}}  \left[ \check{H}^{\rm N 0}(\Vec{R}) + \check{\Delta}(\Vec{R},\Vec{k}) + \check{\Sigma}(\Vec{R},\Vec{k},i \omega_{n})\right] \cdot \Vec{\nabla}_{\Vec{k}} \check{G}(\Vec{R},\Vec{k},i \omega_{n})
 + \frac{i}{2 } 
 \Vec{\nabla}_{\Vec{k}}  \check{H}^{\rm N 1}(\Vec{k})  \cdot \Vec{\nabla}_{\Vec{R}} \check{G}(\Vec{R},\Vec{k},i \omega_{n})
 = \check{1}. \label{eq:wig1}
\end{align}
\end{widetext}
The above equations are simultaneous differential equations with $\Vec{k}$ and $\Vec{R}$. 
In a single-band system, the above equations becomes the differential equations with respect to $\Vec{R}$ with a parameter $\Vec{k}_{\rm F}$ (i.e. the quasiclassical Eilenberger equations), 
since we can eliminate the $\Vec{k}$-mixing term $\Vec{\nabla}_{\Vec{k}} \check{G}(\Vec{R},\Vec{k},i \omega_{n})$, with the use of the contour integration with respect to the energy $\xi_{\Vec{k}}$ and the relation $\Vec{\nabla}_{\Vec{k}} \propto \partial/\partial \xi_{\Vec{k}}$. 
Thus, in order to derive the multi-band quasiclassical Eilenberger equations, 
the $\Vec{k}$-mixing term $\Vec{\nabla}_{\Vec{k}} \check{G}(\Vec{R},\Vec{k},i \omega_{n})$ has to be eliminated.  
However, we can not simply integrate the above equations with respect to $\xi_{\Vec{k}}$, 
since $\xi_{\Vec{k}}$ is the energy obtained by diagonalizing the Hamiltonian $\hat{H}^{\rm N 1}(\Vec{k})$ 
and the Gor'kov equations are determined in the orbital basis. 
One has to characterize the equation by the Fermi wave momentum in the low-energy region.

%%%
\subsection{Projection to the effective low-energy model}
We introduce the projection matrix to develop the multi-band quasiclassical theory. 
The projection matrix eliminates the degree of freedom about the high energy region. 
We define the $2 N \times 2 N$ projection matrix $\check{P}_{\Vec{k}}$ as 
%%%
\begin{align}
\check{P}_{\Vec{k}} &\equiv U_{\Vec{k}}^{M} U_{\Vec{k}}^{M \dagger}. \label{eq:projection}
\end{align}
Here, the $2 N \times 2 M$ matrix $U_{\Vec{k}}^{M}$ is defined by 
\begin{align}
U_{\Vec{k}}^{M}&\equiv 
\left(\begin{array}{cc}
\tilde{U}_{\Vec{k}}^{M} & 0 \\
0 & \tilde{U}_{-\Vec{k}}^{M \ast}
%0 & \tilde{U}_{-\Vec{k}}^{M \ast}
\end{array}\right), \label{eq:matu}
\end{align}
where the $N \times M$ matrix is defined by Eq.~(\ref{eq:ukm}). 
The projection operator satisfies the relation 
\begin{align}
\check{P}_{\Vec{k}}\check{P}_{\Vec{k}} &= \check{P}_{\Vec{k}},
\end{align}
since the matrix $U_{\Vec{k}}^{M}$ always satisfies the relation 
\begin{align}
U_{\Vec{k}}^{M \dagger} U_{\Vec{k}}^{M} &= \bar{1}_{2 M \times 2 M}.  \label{eq:uu}
\end{align}

In order to show the physical meaning of the projection matrix, we operate $\check{P}_{\Vec{k}}$ on the homogeneous $N$-band Green's function in normal states determined by 
\begin{align}
\check{G}(\Vec{k}, i \omega_{n}) &= (i \omega_{n}\check{1} - \check{H}^{\rm N 1}(\Vec{k}))^{-1}. 
\end{align}
The component of $\check{G}(\Vec{k}, i \omega_{n})$ is 
expressed as 
\begin{align}
G_{\alpha \beta}(\Vec{k}, i \omega_{n}) &= 
\sum_{\gamma=1}^{N} 
\frac{\tilde{U}_{\alpha \gamma}(\Vec{k}) \tilde{U}_{\beta \gamma}^{\ast}(\Vec{k})}
{i \omega_{n} - \epsilon_{\gamma}(\Vec{k})}, &
  (1 \leq \alpha,\beta \leq N) 
\end{align}
Here, the $N \times N$ unitary matrix $\tilde{U}(\Vec{k})$ diagonalizes $\hat{H}^{\rm N 1}(\Vec{k})$ 
and $\epsilon_{\gamma}(\Vec{k})$ is the $\gamma$-th eigenvalue of $\hat{H}^{\rm N 1}(\Vec{k})$. 
Operating $\check{P}_{\Vec{k}}$ on $\check{G}(\Vec{k}, i \omega_{n})$, we obtain 
\begin{align}
\left[ \check{P}_{\Vec{k}} \check{G}(\Vec{k}, i \omega_{n})\right]_{\alpha \beta} &=
\sum_{\gamma=1}^{M} 
\frac{\tilde{U}_{\Vec{k}\alpha \gamma}^{M}\tilde{U}_{\Vec{k} \beta \gamma}^{M \ast}}
{i \omega_{n} - \xi_{\Vec{k}}}, 
\end{align}
with $1 \leq \alpha,\beta \leq N$. 
Here, we use the assumption in (\ref{eq:eigen}). 
The above equation means that the projection operator $\check{P}_{\Vec{k}}$ eliminates 
the information of large eigenvalues from $\check{G}(\Vec{k}, i \omega_{n})$. 
The difference $\delta \check{G}(\Vec{k}, i \omega_{n}) \equiv \check{G}(\Vec{k}, i \omega_{n})- \check{P}_{\Vec{k}} \check{G}(\Vec{k}, i \omega_{n})$ 
becomes 
\begin{align}
\left[ \delta \check{G}(\Vec{k}, i \omega_{n}) \right]_{\alpha \beta}&=  
 \sum_{\gamma, \epsilon_{\gamma} \neq 0}^{N-M} 
\frac{\tilde{U}_{\alpha \gamma}(\Vec{k}) \tilde{U}_{\beta \gamma}^{\ast}(\Vec{k})}
{i \omega_{n} - \epsilon_{\gamma}(\Vec{k})} , 
\end{align}
with $1 \leq \alpha,\beta \leq N$. 
If the eigenvalues are located far from the Fermi energy $(|\omega_{n}| \ll \epsilon_{\gamma}(\Vec{k}_{\rm F}))$,   
$\delta \check{G}(\Vec{k}_{\rm F}, i \omega_{n})$ becomes negligible small so that 
we can obtain the relation  expressed as 
\begin{align}
\check{P}_{\Vec{k}} \check{G}(\Vec{k}, i \omega_{n}) &\sim   \check{G}(\Vec{k}, i \omega_{n}), \label{eq:pg}
\end{align}
which is appropriate in the low-energy region. 

We introduce the $2 M \times 2 M$ reduced matrix $\bar{A}$  
expressed as 
\begin{align}
\bar{A}(\Vec{k})  &\equiv U_{\Vec{k}}^{M \dagger} \check{A}(\Vec{k}) U_{\Vec{k}}^{M},  
\end{align}
where $\check{A}$ is the $2 N \times 2 N$ matrix used in the Gor'kov equation (\ref{eq:wig}). 
With the use of Eqs.~(\ref{eq:uu}) and (\ref{eq:pg}), 
$\check{A}$ can be expressed by 
\begin{align}
\check{A}(\Vec{k})  &\sim U_{\Vec{k}}^{M} \bar{A}(\Vec{k})  U_{\Vec{k}}^{M \dagger}, \label{eq:areduced}
\end{align}
in the low-energy region ($\Vec{k} \sim \Vec{k}_{\rm F}$).

\subsection{Multi-band quasiclassical Green's function}
Let us construct the $2 M \times 2 M$ 
quasiclassical multi-band Eilenberger equations in the projected space. 
By multiplying the the both sides in Eq.~(\ref{eq:wig1}) by the matrices $U_{\Vec{k}}^{M \dagger}$ and $U_{\Vec{k}}^{M}$, subtracting the right-hand Gor'kov equation,  and integrating over $\xi_{\Vec{k}}$ (in detail, see Appendix \ref{sec:quasieilen}), 
we obtain $2 M \times 2 M$ quasiclassical multi-band Eilenberger equation 
expressed as 
\begin{widetext}
\begin{align}
 i \Vec{v}_{\rm F}(\Vec{k}_{\rm F}) \cdot \Vec{\nabla}_{\Vec{R}}
  \bar{g}_{\Vec{R}}(\Vec{k}_{\rm F},z)   +
  \left[
 z  \bar{\sigma}_{z}-
   \bar{V}_{0 \Vec{R}}(\Vec{k}_{\rm F}) \bar{\sigma}_{z} 
   -
   \bar{\Delta}_{\Vec{R}}(\Vec{k}_{\rm F})\bar{\sigma}_{z}  -
\bar{\Sigma}_{\Vec{R}} (\Vec{k}_{\rm F},z) \bar{\sigma}_{z} 
,
 \bar{g}_{\Vec{R}}(\Vec{k}_{\rm F},z) 
   \right]_{-}
&= 0, \label{eq:multi}
\end{align}
\end{widetext}
where we introduce the $2 M \times 2 M$ Green's function, non-local potentials, order parameters, and self-energies determined by 
%%%%%
\begin{align}
\bar{G}_{\Vec{R}} (\Vec{k},z) 
&\equiv U_{\Vec{k}}^{M \dagger} \check{G} (\Vec{R},\Vec{k},z) U_{\Vec{k}}^{M}  \label{eq:greenu} \\
\bar{V}_{0 \Vec{R}}(\Vec{k}) &\equiv   U_{\Vec{k}}^{M \dagger} \check{H}^{\rm N 0}(\Vec{R})U_{\Vec{k}}^{M}, \label{eq:vu}\\
\bar{\Delta}_{\Vec{R}}(\Vec{k}) &\equiv  U_{\Vec{k}}^{M \dagger}
\check{\Delta}(\Vec{R},\Vec{k})
U_{\Vec{k}}^{M}, \label{eq:deltau} \\
\bar{\Sigma}_{\Vec{R}}(\Vec{k}, z) &\equiv  U_{\Vec{k}}^{M \dagger}
\check{\Sigma}(\Vec{R},\Vec{k}, z)
U_{\Vec{k}}^{M}. \label{eq:sigmau}
\end{align}
%%%%%
Here, %\begin{widetext}
$\bar{\sigma}_{z}$ denotes the Pauli matrix in the Nambu-Gor'kov space and 
we define the $2 M \times 2 M$ quasiclassical Green's function expressed as 
\begin{align}
\bar{g}_{\Vec{R}}(\hat{\Vec{k}}_{\rm F},z) &\equiv \oint d\xi_{\Vec{k}}\bar{\sigma}_{z} \bar{G} (\Vec{R},\Vec{k},z),  \label{eq:quasi}
\end{align}
where $\oint$ means taking the contributions from poles close to the Fermi surface. 
The matrix structure of the above equation is equivalent to that of the single-band Eilenberger equation. 
When the eigenvalue is not degenerated ($M = 1$), the $2 \times 2$ equation (\ref{eq:multi}) can be regarded as that in a spin-singlet single band superconductor\cite{Serene,Choi,NagaiJPSJ}.   
Therefore, we call the $2 M \times 2 M$ matrix representation the ``single-band description''.
The band index $\alpha$ is useful to compare with the present multi-band theory by 
replacing $\bar{g}_{\Vec{R}}^{\alpha}(\hat{\Vec{k}}_{\rm F},z)$ as $\bar{g}_{\Vec{R}}(\hat{\Vec{k}}_{\rm F},z)$. 
The multi-band Eilenberger equations (\ref{eq:multi}) are similar to the decoupled Eilenberger equations Eq.~(\ref{eq:mgb2type}).
We should note that our multi-band theory includes the orbital characters, since all matrices are determined in the projected space. 
For example, the self-energy with the $T$-matrix approximation depends on the momentum as discussed in Eq.~(\ref{eq:sigma2}).

%%%%

We should note that the normalization condition is equivalent to that in the single-band system expressed as (in detail, see Appendix \ref{sec:normalization})
\begin{align}
\bar{g} \bar{g} &= - \pi^{2} \bar{1}. \label{eq:normalization}
\end{align}

%%%%%%%
%%%%%%
\subsection{Relations in the projected space}
Let us discuss the relations satisfied in the projected space. 
$M \times M$ order parameter is defined in Eq.~(\ref{eq:gapeff}).
If the original order parameter have the relation $\hat{\Delta}(\Vec{R},\Vec{k})^{\dagger} \hat{\Delta}(\Vec{R},\Vec{k}) = \Delta_{0}(\Vec{R},\Vec{k}) 1_{N \times N}$, 
the projected order parameter have the similar relation:  
\begin{align}
\Delta_{\rm eff}(\Vec{R},\Vec{k})^{\dagger} \Delta_{\rm eff}(\Vec{R},\Vec{k}) &\sim \Delta_{0}(\Vec{R},\Vec{k}) 1_{M \times M}, 
\end{align}
because of $\tilde{U}_{\Vec{k}}^{M } \tilde{U}_{\Vec{k}}^{M \dagger} \hat{\Delta}(\Vec{R},\Vec{k})   \sim \hat{\Delta}(\Vec{R},\Vec{k}) $ near the Fermi energy. 
Here, $\Delta_{\rm eff}(\Vec{R},\Vec{k})$ is determined in Eq.~(\ref{eq:gapeff}). 
This indicates that the unitarity of the order parameter is conserved in the projected space.

%
%
%
%The Riccati amplitudes $\bar{a}$ and $\bar{b}$ 

%%%
\subsection{Physical quantities and Gap equations}
Let us express physical quantities with the use of the multi-band quasiclassical Green's function. 
By substituting Eq.~(\ref{eq:areduced}) into Eq.~(\ref{eq:wigldos}), 
the local density of states is expressed as 
\begin{align}
N(\Vec{r},E) &= \frac{-1}{\pi} {\rm Im} \: \left[ \lim_{\eta \rightarrow +0} \sum_{\Vec{k}} {\rm Tr} \: 
\bar{G}^{11}(\Vec{r},\Vec{k},i \omega_{n} \rightarrow E + i \eta) \right] , \\
&= \frac{-1}{\pi} {\rm Im} \: \left[  \lim_{\eta \rightarrow +0} \langle  {\rm Tr} \:  
\bar{g}^{11}(\Vec{r},\Vec{k},i \omega_{n} \rightarrow E + i \eta) \rangle_{\hat{\Vec{k}}_{\rm F}} \right],
\end{align}
where $\bar{G}^{11}$ and $\bar{g}^{11}$ denote $(1,1)$-element in the particle-hole space,  
the bracket denotes the Fermi-surface average $\langle \cdots \rangle_{\hat{\Vec{k}}_{\rm F}} = \int \cdots d S_{\rm F}(\hat{\Vec{k}}_{\rm F})|\Vec{v}_{\rm F}(\hat{\Vec{k}}_{\rm F})|^{-1}/\int d S_{\rm F}(\hat{\Vec{k}}_{\rm F})|\Vec{v}_{\rm F}(\hat{\Vec{k}}_{\rm F})|^{-1}$. 
Other physical quantities can be expressed by the multi-band quasiclassical Green's function in the same manner. 

Finally, we complete the multi-band quasiclassical theory by giving the self-consistent equations for the order parameters.
The gap equations in the Wigner representation is given as 
\begin{align}
\hat{\Delta}_{ \alpha \beta} (\Vec{R},\Vec{k}) &= - T 
\sum_{\Vec{k}', \gamma \gamma'}^{N} \sum_{n} V_{\beta \alpha ; \gamma \gamma'}(\Vec{k},\Vec{k}') 
\hat{F}_{\gamma \gamma'}(\Vec{R},\Vec{k}', i \omega_{n})
\end{align}
By using Eq.~(\ref{eq:deltau}), 
the $M \times M$ order parameter matrix is expressed by
\begin{align}
&\Delta_{{\rm eff} l_{1} l_{2} \Vec{R}}(\Vec{k}_{\rm F}) = \nonumber \\
&
-  T \sum_{n} \sum_{l_{3} l_{4}}^{M} %\nonumber \\
%&\times 
\langle
\bar{V}^{l_{1} l_{2}}_{l_{3} l_{4}}(\hat{\Vec{k}}_{\rm F},\hat{\Vec{k}}_{\rm F}')
\bar{g}^{12}_{l_{3} l_{4} \Vec{R}}( \hat{\Vec{k}}_{\rm F}',i \omega_{n})
 \rangle_{\hat{\Vec{k}}_{\rm F}'}, 
\end{align}
where $\bar{g}^{12}$ denotes $(1,2)$-element in the particle-hole space 
 and  
$\bar{V}^{l_{1} l_{2}}_{l_{3} l_{4}}(\hat{\Vec{k}}_{\rm F},\hat{\Vec{k}}_{\rm F}')$ is the effective interaction expressed as 
%%%%
\begin{align}
&\bar{V}^{l_{1} l_{2}}_{l_{3} l_{4}}(\hat{\Vec{k}}_{\rm F},\hat{\Vec{k}}_{\rm F}') \equiv
\sum_{\alpha \beta \gamma \gamma'} 
V_{\beta \alpha ; \gamma \gamma'}(\Vec{k}_{\rm F},\Vec{k}_{\rm F}') \nonumber \\
&\times 
\tilde{U}_{\Vec{k}_{\rm F} \alpha l_{1}}^{M \ast} \tilde{U}_{-\Vec{k}_{\rm F} \beta l_{2}}^{M \ast}
 \tilde{U}_{\Vec{k}_{\rm F}' \gamma l_{3}}^{M} \tilde{U}_{-\Vec{k}_{\rm F}' \gamma' l_{4}}^{M \ast}.
\end{align}
We can simplify the above gap equations if the pairing interaction has a separable form expressed as\cite{Allen} 
\begin{align}
V_{\beta \alpha ; \gamma \gamma'}(\Vec{k},\Vec{k}') &= 
V_{\alpha \beta}(\Vec{k}) V_{\gamma \gamma'}(\Vec{k}'). 
\end{align}
Here, we use the relation $V_{\beta \alpha ; \gamma \gamma'}(\Vec{k},\Vec{k}') = 
V_{\gamma' \gamma ; \alpha \beta}(\Vec{k}',\Vec{k})$\cite{SigristUeda}. 
The gap equations in this case are given as 
\begin{align}
\Delta_{\Vec{R}} &= 
-  T \sum_{n} \sum_{l_{3} l_{4}}^{M} %\nonumber \\
%&\times 
\langle
\tilde{V}_{l_{3} l_{4}}^{\ast}(\hat{\Vec{k}}_{\rm F}')
\bar{g}^{12}_{l_{3} l_{4} \Vec{R}}( \hat{\Vec{k}}_{\rm F}',i \omega_{n})
 \rangle_{\hat{\Vec{k}}_{\rm F}'}, 
\end{align}
with 
\begin{align}
\Delta_{{\rm eff}\Vec{R}}(\Vec{k}_{\rm F}) &\equiv \Delta_{\Vec{R}}  \tilde{V}(\hat{\Vec{k}}_{\rm F}), \\
\tilde{V}(\hat{\Vec{k}}_{\rm F}) &\equiv  \tilde{U}_{\Vec{k}_{\rm F}}^{M \dagger} V(\Vec{k}_{\rm F}) 
\tilde{U}_{-\Vec{k}_{\rm F}}^{M \ast},  
\end{align}
where we assume that $V_{\gamma \gamma'}(\Vec{k}')^{\ast} = V_{\gamma \gamma'}(\Vec{k}')$. 

\subsection{Perturbative approach in the quasiclassical theory: Zeeman and spin-orbit couplings}
In this section, we discuss the method to treat the Zeeman and spin-orbit couplings. 
In the previous studies\cite{Alexander:1985jh,Vorontsov:2010it,Ichioka:2007dp,Hayashi,Rieck}, 
they used the $4 \times 4$ matrix quasiclassical Eilenberger equations in spin and Nambu spaces expressed as 
\begin{align}
&i \Vec{v}_{\rm F} \cdot \Vec{\nabla}_{\Vec{R}} \check{g}_{\Vec{R}}(\Vec{k}_{\rm F},z) \nonumber \\
&+\left[
z \sigma_{z} -\check{\Delta}\sigma_{z} - \check{H}_{1}(\Vec{k}_{\rm F}) \sigma_{z},\check{g}_{\Vec{R}}(\Vec{k}_{\rm F},z)
 \right]_{-} = 0.
\end{align}
Here, $\sigma_{i}$ denotes the Pauli matrix in the Nambu space and $\check{H}_{1}(\Vec{k}_{\rm F})$ 
includes the spin-orbit and/or Zeeman coupling terms.  
These quasiclassical equations can describe a vortex state in a Fulde-Ferrell-Larkin-Ovchinnikov superconductor\cite{Ichioka:2007dp}. 
In terms of our theory, the above equations are obtained by assuming that the number of degenerated Fermi surfaces is two ($M = 2$). 
In general, however, the Zeeman and spin-orbit interactions split the degenerated bands (i.~e.~ $M = 2 \rightarrow M = 1$ ). 
Thus, it is found that the previous studies assume that the Zeeman and spin-orbit interactions are weak. 
With the use of the perturbative approach in the multi-band quasiclassical theory, we can derive the above equations as follows.  
Let us divide $\hat{H}^{\rm N1}(\Vec{k})$ in Eq.~(\ref{eq:eigen}) into the two terms as 
\begin{align}
\hat{H}^{\rm N1}(\Vec{k}) &= \hat{H}^{\rm N1}_{0}(\Vec{k}) + \hat{H}_{1}(\Vec{k}). 
\end{align}
Thus, in order to construct the projection operator $\check{P}_{\Vec{k}}$, we can use the eigenvectors obtained by 
\begin{align}
\hat{H}^{\rm N1}_{0}(\Vec{k})  u_{\Vec{k}}^{i} &= \xi_{\Vec{k}} u_{\Vec{k}}^{i}. 
\end{align}
This approach is appropriate for the case that the inter-band pairing between the different Fermi surfaces is important. 
The perturbative Zeeman field enables us to treat the Pauli paramagnetic depairing in the quasiclassical framework.

\subsection{Riccati-type equations}
It is known that it is difficult to numerically solve the quasiclassical Eilenberger equations\cite{NagaiMeso}, since 
the equations have a divergent solution as a particular solution. 
A careful computational treatment is required for integrating the Eilenberger equations with the use of  the so-called explosion method\cite{Thuneberg:1982fp}. 
To avoid this difficulty, the Riccati-type equations, which are obtained by a special parametrization form of the quasiclassical Green's function, are used\cite{YKato,Nagato,Higashitani,NagatoLow,SchopohlMaki,Schopohl}. 
In addition, to solve the Riccati equation stably, we have proposed the efficient numerical method for obtaining unique solutions in the single-band Eilenberger 
framework\cite{NagaiMeso}.  
We show that it is easy to expand this method into the multi-band systems. 
For simplicity, we neglect the self-energy $\bar{\Sigma} = 0$. 
We use a special parametrization form of the quasiclassical Green's function to solve Eq.~(\ref{eq:multi}). 
The solution $\bar{g}$ of Eq.~(\ref{eq:multi}) can be written as, 
\begin{align}
\bar{g} &= - i \pi \bar{N} \left(\begin{array}{cc}
(\tilde{1} - \tilde{a} \tilde{b}) & 2 i \tilde{a} \\
-2 i \tilde{b} & -(\tilde{1}- \tilde{b} \tilde{a})
\end{array}\right), \\
\bar{N} &= \left(\begin{array}{cc}
(\tilde{1} + \tilde{a} \tilde{b})^{-1} & 0 \\
0 & (\tilde{1}+ \tilde{b} \tilde{a})^{-1}
\end{array}\right), 
\end{align}
where $M \times M$ matrices $\tilde{a}$ and $\tilde{b}$ are the solutions of the 
following matrix-type Riccati differential equations:
\begin{align}
\Vec{v}_{\rm F} \cdot \Vec{\nabla} \tilde{a} &= - 2 \omega_{n} \tilde{a} - \tilde{a} \tilde{\Delta}^{\dagger} \tilde{a} + \tilde{\Delta}. \\
\Vec{v}_{\rm F} \cdot \Vec{\nabla} \tilde{b} &=   2 \omega_{n} \tilde{b} + \tilde{b} \tilde{\Delta} \tilde{b} - \tilde{\Delta}^{\dagger}.
\end{align}
Since the above equations contain $\Vec{\nabla}$ only through $\Vec{v}_{\rm F} \cdot \Vec{\nabla}$, 
these can be reduced to a one-dimensional problem on a straight line in the direction of the 
Fermi velocity $\Vec{v}_{\rm F}$:
\begin{align}
v_{\rm F} \frac{\partial \tilde{a} }{\partial s} &= - 2 \omega_{n} \tilde{a} - \tilde{a} \tilde{\Delta}^{\dagger} \tilde{a} + \tilde{\Delta}. \\
v_{\rm F}  \frac{\partial \tilde{b} }{\partial s} &=   2 \omega_{n} \tilde{b} + \tilde{b} \tilde{\Delta} \tilde{b} - \tilde{\Delta}^{\dagger}.
\end{align}
In a bulk system with $\tilde{\Delta}^{\dagger} \tilde{\Delta} \propto \tilde{1}$, the solutions of the Riccati equations are 
\begin{align}
\tilde{a}(\omega_{n}) &= \frac{\tilde{\Delta}}{\omega_{n} + \sqrt{\omega_{n}^{2} + \frac{1}{2} {\rm Tr} \: \tilde{\Delta} \tilde{\Delta}^{\dagger}}}, \label{eq:riccatia}\\
\tilde{b}(\omega_{n}) &= \frac{\tilde{\Delta}^{\dagger}}{\omega_{n} + \sqrt{\omega_{n}^{2} + \frac{1}{2} {\rm Tr} \: \tilde{\Delta} \tilde{\Delta}^{\dagger}}}. \label{eq:riccatib}
\end{align}
According to the previous paper\cite{NagaiMeso}, putting $\tilde{a}(s_{a}) = 0$ and $\tilde{b}(s_{b}) =0$ as initial values, 
one can obtain physical solutions by integrating the Riccati equations.
Here, $s_{a}$ and $s_{b}$ are initial spatial points.

%%%%%%%%%
\subsection{Boundary condition for the Riccati parameters at a specular surface}
To solve Riccati-type differential equations, one has to consider the boundary condition for Riccati parameters $\tilde{a}$ and $\tilde{b}$. 
In this section, we consider a specular surface. 
In the case of $M = 1$ or $M = 2$, we can show the transformation of the linearized BdG equations to the matrix 
Riccati equations. 
Thus, the boundary condition for the Riccati parameters can be determined explicitly. 
We note that, in many materials, the number of the degenerated Fermi levels $M$ is not larger than $M = 2$. 

If the Fermi level with a certain momentum $\Vec{k}_{\rm F}$ is not degenerated ($M = 1$), the Riccati equations are derived by the relation: 
\begin{align}
\tilde{a}(\Vec{r},\Vec{k}_{\rm F}) = i \frac{f^{\Vec{k}_{\rm F}}(\Vec{r})}{g^{\Vec{k}_{\rm F}}(\Vec{r})}.
\end{align}
With the use of the boundary condition for wavefunctions in Eqs.~(\ref{eq:fgbf}) and (\ref{eq:fgb}), the boundary condition at a surface $z = 0$ with $K = 2$ is given by 
\begin{align}
\tilde{a}(k_{{\rm F}z}^{2}) &=  e^{- 2 i \Phi_{12}} \tilde{a}(k_{{\rm F}z}^{1}).
\end{align}
If the Fermi level is doubly degenerated ($M = 2$), we find the relation expressed as 
\begin{align}
\tilde{a}(\Vec{r},\Vec{k}_{\rm F}) &= i U(\Vec{k}_{\rm F},\Vec{r}) V(\Vec{k}_{\rm F},\Vec{r})^{-1}
\end{align}
where the $2 \times 2$ matrices $U(\Vec{k}_{\rm F},\Vec{r})$ and $V(\Vec{k}_{\rm F},\Vec{r})$ are determined by 
\begin{align}
U(\Vec{k}_{\rm F},\Vec{r}) &\equiv
\left(\begin{array}{cc}
\Vec{f}^{\Vec{k}_{\rm F}}(\Vec{r}) & - \Vec{g}^{-\Vec{k}_{\rm F}}(\Vec{r})^{\ast}
\end{array}\right), \\
V(\Vec{k}_{\rm F},\Vec{r}) &\equiv 
\left(\begin{array}{cc}
\Vec{g}^{\Vec{k}_{\rm F}}(\Vec{r}) &  \Vec{f}^{-\Vec{k}_{\rm F}}(\Vec{r})^{\ast}
\end{array}\right).
\end{align}
We obtain the boundary condition at a surface $z = 0$ with $K = 2$ and $N = M = 2$ expressed as 
\begin{align}
\tilde{a}(k_{{\rm F}z}^{2}) &= \tilde{V}^{(k_{{\rm F}x},k_{{\rm F}y},k_{{\rm F}z}^{2})}_{(k_{{\rm F}x},k_{{\rm F}y},k_{{\rm F}z}^{1})}
\tilde{a}(k_{{\rm F}z}^{1}) \left[ \tilde{V}^{-(k_{{\rm F}x},k_{{\rm F}y},k_{{\rm F}z}^{2}) \ast}_{-(k_{{\rm F}x},k_{{\rm F}y},k_{{\rm F}z}^{1})} \right]^{-1}. \label{eq:ricboundary}
\end{align}
Note that $\tilde{V}^{(k_{{\rm F}x},k_{{\rm F}y},k_{{\rm F}z}^{2})}_{(k_{{\rm F}x},k_{{\rm F}y},k_{{\rm F}z}^{1})}$ is determined in Eq.~(\ref{eq:transfer}).

%%%%%%%
\subsection{Arbitrary transformation about normal-state eigenvectors: Appearance condition of the Andreev bound states}\label{sec:arbit}
We discuss an arbitrariness of the $N \times M$ matrix $\tilde{U}_{\Vec{k}_{\rm F}}^{\rm M}$. 
With the use of this arbitrariness, one can discuss the appearance condition of the Andreev bound states in multi-band superconductors.
As shown in Eq.~(\ref{eq:ukm}), the $N \times M$ matrix $\tilde{U}_{\Vec{k}_{\rm F}}^{\rm M}$ consists of the zero-energy degenerated eigen vectors about 
the normal-state Hamiltonian $\hat{H}^{\rm N 1}(\Vec{k})$. 
Because of the degeneracy, it should be noted that the $N \times M$ matrix $\tilde{U}_{\Vec{k}_{\rm F}}^{\rm M}$ has additional degrees of freedom expressed as 
\begin{align}
\tilde{U}_{\Vec{k}_{\rm F}}^{\rm M'} &= \tilde{U}_{\Vec{k}_{\rm F}}^{\rm M} \hat{A}_{\Vec{k}_{\rm F}}, 
\end{align}
with a $M \times M$ arbitrary unitary matrix $\hat{A}_{\Vec{k}_{\rm F}}$. 
Although this matrix does not change the $2 N \times 2 M$ projection matrix $\check{P}_{\Vec{k}_{\rm F}}$, 
the representation of the effective order parameters $\Delta_{\rm eff}(\Vec{r},\Vec{k}_{\rm F})$ determined in 
Eq.~(\ref{eq:gapeff}) depends on the matrix $\hat{A}_{\Vec{k}_{\rm F}}$, expressed as 
\begin{align}
\Delta_{\rm eff}(\Vec{r},\Vec{k}_{\rm F}) &= \hat{A}^{\dagger}_{\Vec{k}_{\rm F}} \tilde{U}_{\Vec{k}_{\rm F}}^{M \dagger} \hat{\Delta}(\Vec{r},\Vec{k}_{\rm F})  \tilde{U}_{-\Vec{k}_{\rm F}}^{M \ast} \hat{A}^{\ast}_{-\Vec{k}_{\rm F}}.
\end{align}
It should be noted that this arbitrary transformation does not change any physical quantities, since the matrix  $\hat{A}_{\Vec{k}_{\rm F}}$ changes the boundary condition. 
With the use of the unitary matrix $\hat{A}_{\Vec{k}_{\rm F}}$, we can simplify the boundary condition as follows. 

In the case of $K = 2$ and $N = M$, the boundary conditions (\ref{eq:fnm}) and (\ref{eq:gnm}) become 
\begin{align}
 \Vec{f}^{k_{z}^{2}}(z=0) &= -  \tilde{V'}^{(\Vec{k}_{{\rm F} \parallel},k_{z}^{2})}_{(\Vec{k}_{{\rm F} \parallel},k_{z}^{1})} 
 \Vec{f}^{k_{z}^{1}}(z=0),  \\
 \Vec{g}^{k_{z}^{2}}(z=0)  &= -\tilde{V'}^{(-\Vec{k}_{{\rm F} \parallel},-k_{z}^{2}) \ast}_{(-\Vec{k}_{{\rm F} \parallel},-k_{z}^{1})} 
  \Vec{g}^{k_{z}^{1}}(z=0),
\end{align}
where 
\begin{align}
 \tilde{V'}^{(\Vec{k}_{{\rm F} \parallel},k_{z}^{2})}_{(\Vec{k}_{{\rm F} \parallel},k_{z}^{1})} &\equiv 
\hat{A}_{(\Vec{k}_{{\rm F} \parallel},k_{z}^{2})} \tilde{V}^{(\Vec{k}_{{\rm F} \parallel},k_{z}^{2})}_{(\Vec{k}_{{\rm F} \parallel},k_{z}^{1})} 
 \hat{A}_{(\Vec{k}_{{\rm F} \parallel},k_{z}^{1})}^{\dagger}.
\end{align}
By using the matrix $\hat{A}_{\Vec{k}_{\rm F}}$ satisfying the relation 
\begin{align}
\hat{A}_{(\Vec{k}_{{\rm F} \parallel},k_{z}^{2})} &=  \hat{A}_{(\Vec{k}_{{\rm F} \parallel},k_{z}^{1})}  (\tilde{V}^{(\Vec{k}_{{\rm F} \parallel},k_{z}^{2})}_{(\Vec{k}_{{\rm F} \parallel},k_{z}^{1})} )^{\dagger}, 
\end{align}
we obtain the simplified boundary condition expressed as 
\begin{align}
 \Vec{f}^{k_{z}^{2}}(z=0) &= -  
 \Vec{f}^{k_{z}^{1}}(z=0),  \\
 \Vec{g}^{k_{z}^{2}}(z=0)  &= -
  \Vec{g}^{k_{z}^{1}}(z=0).
\end{align}
In addition, in the case of $K = M = N = 2$, the boundary condition for the Riccati amplitudes (\ref{eq:ricboundary}) becomes 
\begin{align}
\tilde{a}(k_{{\rm F}z}^{2}) &= \tilde{a}(k_{{\rm F}z}^{1}),  
\end{align}
The above boundary condition is equivalent to that for spin-triplet superconductivity in  the past quasiclassical treatment. 
In the case of an effective one-band system ($M = 1$) with $K = 2$, the $1 \times 1$ unitary matrix $\hat{A}_{\Vec{k}_{\rm F}}$ is rewritten as $\hat{A}_{\Vec{k}_{\rm F}} = e^{i \phi_{\Vec{k}_{\rm F}}}$. 
Thus, we can erase the overall phase $\Phi_{12}$ in Eq.~(\ref{eq:k2m1}), 
The boundary condition becomes 
\begin{align}
f^{k_{z}^{2}}(z=0) &= - 
f^{k_{z}^{1}}(z=0), \\
g^{k_{z}^{2}}(z=0) &= - g^{k_{z}^{1}}(z=0).  
\end{align}
We obtain the boundary condition for the Riccati amplitude expressed as 
\begin{align}
\tilde{a}(k_{{\rm F}z}^{2}) &= \tilde{a}(k_{{\rm F}z}^{1}), \label{eq:m1r}
\end{align}
which is completely equivalent to the boundary condition in the single-band quasiclassical Eilenberger theory. 

Now, we can discuss the appearance condition of the Andreev bound state at a surface in multi-band superconductors. 
In a single band model, the Andreev bound states occur when the sign of the gap function changes through the scattering process. 
In the case of $K = 2$ and $M = 1$, we can easily discuss the appearance condition of the Andreev bound states. 
Note that the bound states appear if the condition (\ref{eq:u12}) is satisfied in this case. 
To use the above boundary conditions (\ref{eq:m1r}), the order parameter matrix after the scattering process should be 
\begin{align}
\Delta_{\rm eff}(\Vec{r},\Vec{k}_{{\rm F}}^{2}) &= 
%e^{\left[
e^{ - i \left( \Phi^{\Vec{k}_{{\rm F}}^{2}}_{\Vec{k}_{{\rm F}}^{1}}
+ \Phi^{-\Vec{k}_{{\rm F}}^{2}}_{-\Vec{k}_{{\rm F}}^{1}}
\right)
}
\tilde{U}_{\Vec{k}_{\rm F}^{2}}^{M \dagger} \hat{\Delta}(\Vec{r},\Vec{k}_{\rm F}^{2})  \tilde{U}_{-\Vec{k}_{\rm F}^{2}}^{M \ast}
%\Delta_{\rm eff}(\Vec{r},\Vec{k}_{{\rm F}}^{2}) 
%\exp \left[ - i \Phi^{-(\Vec{k}_{{\rm F} \parallel},k_{{\rm F}z}^{2})}_{-(\Vec{k}_{{\rm F} \parallel},k_{{\rm F}z}^{1})} \right]
\end{align}%
with 
\begin{align}
 e^{i \Phi^{\Vec{k}_{{\rm F}}^{2}}_{\Vec{k}_{{\rm F}}^{1}}
}
&\equiv 
 \tilde{U}_{\Vec{k}_{{\rm F}}^{2}}^{M \dagger}  \tilde{U}_{\Vec{k}_{{\rm F}}^{1}}^{M }.
 \end{align}
 The quasiclassical Green's function at a surface diverges when the relation 
\begin{align}
1 + a(i \omega_{n} \rightarrow \epsilon + i \eta, \Vec{k}_{{\rm F}}^{1}) b(i \omega_{n} \rightarrow \epsilon + i \eta, \Vec{k}_{{\rm F}}^{2}) &= 0
\end{align}
is satisfied.\cite{NagaiJPSJ}
With the use of the bulk solutions in Eqs.~(\ref{eq:riccatia}) and (\ref{eq:riccatib}), 
the appearance condition of the zero-energy Andreev bound states becomes
\begin{align}
|\Delta_{\rm eff}(\Vec{k}_{{\rm F}}^{1})| |\Delta_{\rm eff}(\Vec{k}_{{\rm F}}^{2})| + \Delta_{\rm eff}(\Vec{k}_{{\rm F}}^{1}) \Delta_{\rm eff}(\Vec{k}_{{\rm F}}^{2})^{\ast} = 0.
\end{align}
Thus, the sign of the order parameter is important for the appearance condition of the Andreev bound state even in multi-band superconductors.

\section{Multi-band effects}
\label{sec:multi}
We discuss the physical meanings of the multi-band quasiclassical theory described by Eq.~(\ref{eq:multi}). 
In our theory, the ``multi-band'' effect is characterized by two factors.  
The first factor is how many solutions are in Eq.~(\ref{eq:eigen}) (i.e. the information of the eigenvalues). 
The second one is how the orbitals are mixed in Eq.~(\ref{eq:eigen})  (i.e. the information of the eigenvectors). 

\subsection{Eigenvalues}
\begin{figure}[tb]
\begin{center}
\resizebox{0.7 \columnwidth}{!}{\includegraphics{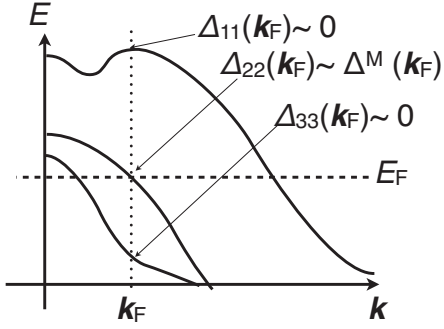}}
\end{center}
\caption{\label{fig:moshiki}The schematic figure of the electron bands in the three-orbital superconductors. 
}
\end{figure}

We discuss the eigenvalues obtained in Eq.~(\ref{eq:eigen}). 
In the quasiclassical theory, the number of the solutions at the Fermi surface $M$ characterizes the multi-band effect. 
For example, in the three-orbital spin-singlet superconductors, the superconducting order parameter is described by 
a $3 \times 3$ matrix ($N = 3$) in Eq.~(\ref{eq:gapeq}). 
Let us consider that the only one band crosses the Fermi energy at the Fermi surface with the Fermi wave number $\Vec{k}_{\rm F}$ ($M = 1$) as shown in Fig.~\ref{fig:moshiki}.
In this case, the other bands must have the much higher or lower energies. 
Thus, the superconducting order parameters on or between these bands (e.~g. $\Delta_{11}(\Vec{k}_{\rm F})$ or $\Delta_{12}(\Vec{k}_{\rm F})$) can not affects the physical quantities, since 
these order parameters ($\sim$ meV) are much smaller than the energy scales of electron bands ($\sim$ eV). 
Therefore, the $1 \times 1$ order parameter matrix on the band crossing the Fermi energy (e.~g. $\Delta_{22}(\Vec{k}_{\rm F})$ in Fig.~\ref{fig:moshiki}) is only effective for the superconductivity. 
In terms of the above point,  
many multi-band superconductors such as MgB$_{2}$ or iron-pnictides can be described by the single-band  superconducting gap because of $M = 1$ in these materials.  
%%%

%%
\subsection{Eigenvectors}
We discuss the eigenvectors in Eq.~(\ref{eq:eigen}), which describes the ratio of the hybridization of the orbital characters at the Fermi wave number $\Vec{k}_{\rm F}$. 
The multi-band effects are described by these eigenvectors. 
For example, we consider the impurity self-energy with the Born approximation. 
%The $2N \times 2N$  matrix self-energy with the Born approximation about random impurities $\check{H}^{\rm N 0}(\Vec{r}) = \sum_{i} \delta(\Vec{r}-\Vec{r}_{i}) \check{V}_{0}$ is described in Eq.~(\ref{eq:bornapr}). 
In our framework, $2 M \times 2 M$  self-energy becomes 
%%%
\begin{align}
& \bar{\Sigma}(\Vec{k}_{\rm F},i \omega_{n}) \bar{\sigma}_{z} =n_{\rm imp}  \bar{V}(\Vec{k}_{\rm F},\Vec{k}_{\rm F})\nonumber \\
& + n_{\rm imp} \langle \bar{V}(\Vec{k}_{\rm F}, \Vec{k}_{\rm F}') \bar{g}(\Vec{k}_{\rm F}',i \omega_{n}) \bar{V}(\Vec{k}_{\rm F}', \Vec{k}_{\rm F}) \rangle_{\Vec{k}_{\rm F}'}, \label{eq:sigma}
\end{align}
%%%
where we determine 
\begin{align}
\bar{V}(\Vec{k}_{\rm F},\Vec{k}_{\rm F}') &\equiv U_{\Vec{k}_{\rm F}}^{M \dagger}
 \check{V}_{0} \check{\sigma}_{z}  U_{\Vec{k}_{\rm F}'}^{M}
.  \label{eq:vmulti}
\end{align} 
In the case of an effective single band system ($M = 1$), $\bar{V}(\Vec{k}_{\rm F},\Vec{k}_{\rm F}') $ becomes the $2 \times 2$ matrix and 
its $(1,1)$-component is expressed as 
\begin{align}
\bar{V}^{11}(\Vec{k}_{\rm F},\Vec{k}_{\rm F}') &=  \Vec{u}_{1}^{\dagger}(\Vec{k}_{\rm F})
 \hat{V}_{0}  \Vec{u}_{1}(\Vec{k}_{\rm F}')
\end{align}
with 
\begin{align}
\hat{H}^{\rm N 1}(\Vec{k}_{\rm F}) \Vec{u}_{1}(\Vec{k}_{\rm F}) &= 0.
\end{align}
In the case of $\hat{V}_{0} \propto \hat{1}$, the strength of the multi-band effects can be determined by the orthogonality between the eigenvectors at the different Fermi momenta.   
We note the case of the $T$-matrix approximation for randomly distributed impurities. 
The $2 N \times 2 N$ matrix self-energy is written as 
\begin{align}
\check{\Sigma}(i \omega_{n})\bar{\sigma}_{z} &= n_{\rm imp} \check{T}(i \omega_{n}),
\end{align}
where
\begin{align}
\check{T}(i \omega_{n}) &= \check{V}_{0} +  \sum_{\Vec{k'}} \check{V}_{0} \check{G}_{\Vec{k}'}(i \omega_{n}) \check{T}(i \omega_{n}).
\end{align}
In our framework, the self-energy with $2 M \times 2 M$ matrix form is obtained as
\begin{align}
\bar{\Sigma}(\Vec{k}_{\rm F}, i \omega_{n}) &= n_{\rm imp} \bar{T}_{\Vec{k}_{\rm F} \Vec{k}_{\rm F}}( i \omega_{n}),
\end{align}
where 
\begin{align}
\bar{T}_{\Vec{k}_{\rm F} \Vec{k}_{\rm F}'}( i \omega_{n}) &= \bar{V}(\Vec{k}_{\rm F},\Vec{k}_{\rm F}') +
 \langle
 \bar{V}(\Vec{k}_{\rm F},\Vec{k}_{\rm F}'')  \bar{g}(\Vec{k}_{\rm F}'',i \omega_{n})  \bar{T}_{\Vec{k}_{\rm F}'',\Vec{k}_{\rm F}'}
 \rangle_{\Vec{k}_{\rm F}''}. \label{eq:selfT}
\end{align}
Thus, the eigenvectors of the normal state Hamiltonian are important to describe the impurity effects.

%%%
\subsection{Non-local anisotropic potentials in the projected space}
As shown in the previous sections, the multi-band Eilenberger equations (\ref{eq:multi}) have the orbital characters through the matrix $\tilde{U}^{M}(\Vec{k}_{\rm F})$. 
It should be noted that the matrix $\tilde{U}^{M}(\Vec{k}_{\rm F})$ makes the potential $\check{H}^{\rm N 0}(\Vec{R})$ {\it non-local} and anisotropic. 
The non-locality originates from the non-unitary transformation which erases the information. 
Non-local potentials have been used as pseudo-potentials in the first-principles calculations. 
In the first-principle calculations, the pseudo-potential method which treats valence electrons only is commonly used 
to erase the degrees of inner-shell electrons. 
In our theory, the non-local potentials are used to erase the fast oscillations characterized by $\Vec{k}_{\rm F}$.  
The multi-band effects are understood by the non-locality and anisotropy in the projected space. 
The effective potential $\bar{V}_{0 \Vec{R}}(\Vec{k}_{\rm F})$ in Eq.~(\ref{eq:multi}) is non-local and anisotropic, 
since the potential depends on the center-of-mass coordinate $\Vec{R}$ and the relative coordinate $\Vec{k}_{\rm F}$.  
The non-locality and anisotropy can be easily understood through the example of the self-energy with the Born equation expressed as Eq.~(\ref{eq:sigma}).
The above self-energy can be regarded as that made from the non-local potential $\bar{V}(\Vec{r},\Vec{r}')$. 
We should note that the non-locality and anisotropy are strong in the iron-based superconductors, since $d$-orbitals are strongly entangled at the Fermi level.   

\subsection{Differences between the present and previous quasiclassical multi-band frameworks}
We discuss the differences between the present and previous quasiclassical multi-band treatments. 
Fundamentally, note that our framework is an extension of a previous single-band Eilenberger framework. 
Thus, we make the derivation similar to the previous single-band one.

Our main point is the low-energy projection which systematically reduces a $N$-band system to a $M$-band system. 
There are two kinds of the reductions in our paper. 
The first one is same as that in the previous paper\cite{Serene}, which uses Fermi velocities and Green's functions at the Fermi level in normal states.
The second one is the reduction of the band-degree of freedom, which was not mentioned in Ref.~{\it et al}. 
Our equations can treat the off-diagonal elements of a Green's function between bands. 
A previous theory treated inter-band effects only through gap-equations not the Eilenberger equations. 
Note that the results in the previous papers using the quasiclassical theory were obtained only in the problem which the previous framework was available. 
Our framework extends the applicable region of the quasiclassical theory.

We claim that there was no multi-band Eilenberger equation which can treat inter-band effects correctly in inhomogeneous systems.
In the previous framework, for example, it is hard to study the vortex bound states with impurities in complicated multi-band superconductors, such as iron-based superconductors. 
One usually considers the five-orbital model for the iron-based superconductors whose Fermi surfaces are constructed by only three-bands. 
In this case, the decoupled Eilenberger equations have three band indices. 
When one introduces a self energy ({\it e.g.} the impurity-induced self energy), the off-diagonal elements of the self energy can not be described by the decoupled Eilenberger equations. 
On the other hand, such a self-energy matrix should be defined by the Dyson equation with Feynman-diagram techniques in five-orbital model. 
In addition, 
band-coupled Eilenberger equations except for our framework can be constructed only in the case that the Fermi velocities are the same in the different bands, 
since the decoupled Eilenberger equations with a band index are characterized by a Fermi velocity on the band.

Finally, we show the example which makes clear difference between our and previous frameworks, qualitatively. 
The corrections in our framework describe the multi-orbital effects. 
Multi-band effects in the present framework are necessary to correctly calculate a reduction of the critical temperature caused by impurities.
While the self energy with T-matrix approximation induced by the randomly-distributed impurities does not depend on the momentum in the 
previous decoupled Eilenberger theory, the self-energy in our Eilenberger equation depends on momentum. 
If the momentum dependence is neglected, the quasiclassical theory can not correctly calculate the reduction of the critical temperature.  
We show the qualitative difference between the present and previous frameworks in the system on the topological insulator in Sec.~\ref{sec:rp}.
The theoretical calculation by directly solving the BdG equations suggested that the proximity-induced superconductivity on the surface of the topological insulator 
is robust against non-magnetic impurities\cite{Ito:2011ct}. 
The present multi-band framework can correctly describe the robustness. 
The previous quasiclassical framework, however, can not reproduce this robustness. 
\section{Multi-band quasiclassical approximations in various kinds of systems: Examples}
\label{sec:single}
In this section, we apply the multi-band quasiclassical theory in the various kinds of systems as examples. 

%%%%%%%%

%%
\subsection{Noncentrosymmetric Superconductors: CePt$_{3}$Si}
\label{sec:CePt}
We show that our multi-band quasiclassical theory makes the past debates clear. 
The noncentrosymmetric superconductor CePt$_{3}$Si has the Rashba-type spin-orbit interaction due to the 
lack of the inversion symmetry\cite{HayashiCe,NagaiCe}. 
The mixed spin-singlet-triplet model has been used to study this material. 
By assuming that the spatial variations of the $s$-wave pairing component of the pair potential are 
the same as those of the $p$-wave pairing component, the gap function is expressed as 
\begin{align}
\hat{\Delta}(\Vec{k}) &= \left[\Psi \sigma_{0} + \Vec{d}_{\Vec{k}} \cdot \Vec{\sigma} \right] i \sigma_{y}. 
\end{align}
Here, $\sigma_{i}$ is the Pauli matrix in spin space. 
The normal-state Hamiltonian with the Rashba-type spin-orbit coupling is written as 
%%%
\begin{align}
\hat{H}(\Vec{k}) &= (\lambda_{\Vec{k}} - \mu) \sigma_{0} + \Vec{g}_{\Vec{k}} \cdot \Vec{\sigma},  \label{eq:ceham}
\end{align}
%%%
where the spin-orbit interaction satisfies the relation $\Vec{g}_{- \Vec{k}} = - \Vec{g}_{\Vec{k}}$. 
Here, $\lambda_{\Vec{k}}$ is the dispersion without the spin-orbit interaction. 
We determine $\Vec{g}_{\Vec{k}}$ as\cite{HayashiCe,NagaiCe} % \equiv (g_{x \Vec{k}},g_{y \Vec{k}},g_{z \Vec{k}})$. 
\begin{align}
\Vec{g}_{\Vec{k}}^{\rm T} &\equiv
g(-\sin \phi \sin \theta, \cos \phi \sin \theta,0).
\end{align}
We assume that the $\Vec{d}$-vector is parallel to $\Vec{g}_{\Vec{k}}^{\rm T} $ expressed as 
\begin{align}
\Vec{d}_{\Vec{k}}^{\rm T} &\equiv \Delta_{d} (-\sin \phi \sin \theta, \cos \phi \sin \theta,0).
\end{align}
In the previous papers\cite{HayashiCe}, from the original Eilenberger equation for noncentrosymmetric superconductivity\cite{HayashiNMR,SchopohlLow}, 
they have obtained two equations corresponding to these split Fermi surface I and II in the case 
of the $s$+$p$-wave pairing state,\cite{HayashiPRB}
\begin{align}
i \Vec{v}_{\rm I,II} \cdot \Vec{\nabla} \check{g}_{\rm I,II} + [i \omega_{n} \check{\tau}_{3} - \check{\Delta}_{\rm I, II},\check{g}_{\rm I, II}]&= 0.  \label{eq:ce}
\end{align}
where $\check{\Delta}_{\rm I, II} = [(\check{\tau}_{1}+ i \check{\tau}_{2})\Delta_{\rm I, II} - (\check{\tau}_{1}- i \check{\tau}_{2} )\Delta_{\rm I,II}^{\ast}]/2$, $\Delta_{\rm I, II} = \psi \pm \Delta_{d} \sin \theta$ are the order parameters on the Fermi surface I and II, $(\check{\tau}_{1},\check{\tau}_{2},\check{\tau}_{3})$ are the Pauli matrices in the particle-hole space, and the commutator $[\check{a},\check{b}]= \check{a}\check{b}- \check{b}\check{a}$. 
There are many successes with the use of the above decoupled equations. 
We should note that there are some debates about the appropriate region of the above approach\cite{Hayashipri}. 
In the real material such as CePt$_{3}$Si, there is the strong spin-orbit coupling ($\sim eV$).
It has been not clear whether this approach is the weak-spin-orbit coupling approach, since the two same-size spherical Fermi surfaces 
are assumed. 
The difference of the size of the each Fermi surface depends on the strength of the spin-orbit coupling. 
On the other hand, the size of the Fermi surfaces is naturally considered in our multi-band theory. 

Let us apply the multi-band quasiclassical theory to the noncentrosymmetric superconductors in order to derive the 
decoupled Eilenberger equations Eq.~(\ref{eq:ce}) directly from the Hamiltonian (\ref{eq:ceham}). 
The eigenvalues of the $2 \times 2$ matrix in Eq.~(\ref{eq:ceham}) is given by 
\begin{align}
\epsilon_{\pm}(\Vec{k}) &= \lambda_{\Vec{k}} - \mu \pm |\Vec{g}_{\Vec{k}}|.
\end{align}
Although there are two Fermi surfaces, the eigenvalue is not degenerated so that we obtain $M = 1$. 
The eigenvectors associated with $\epsilon_{\pm}(\Vec{k})$ are expressed as 
\begin{align}
u_{\Vec{k}}^{+} &=  \frac{1}{\sqrt{2}} \left(\begin{array}{c}
- i e^{- i \phi} \\
1
\end{array}\right),\\
u_{\Vec{k}}^{-} &= \frac{1}{\sqrt{2}}   \left(\begin{array}{c}
1\\
-i e^{i \phi} 
\end{array}\right).
\end{align}
The $1 \times 1$ effective gap function is given as 
\begin{align}
\Delta_{\pm} &= \pm i e^{\pm i \phi} (\Psi \pm  \Delta_{d} \sin \theta).
\end{align}
The above effective gap function is not equivalent to $\Delta_{\rm I,II}$ in the previous paper.  
We should note that a representation of the effective gap function has 
an arbitrary degree of freedom expressed as 
\begin{align}
\Delta_{\rm eff}(\phi,\theta)' &= A(\phi,\theta)^{\dagger} \Delta_{\rm eff}(\phi,\theta) A(\phi+\pi,\theta + \pi)^{\ast}, 
\end{align}
as discussed in Sec.~\ref{sec:arbit}. 
Thus, we can use a $1 \times 1$ arbitrary unitary matrix in order to change 
a representation of the effective gap. 
With the use of the $1 \times 1$ unitary matrix $A_{\pm}(\phi,\theta)$ defined as 
\begin{align}
A_{\pm}(\phi,\theta) &= \pm e^{\pm i \frac{\phi}{2}},  
\end{align}
the effective gap function can be rewritten as 
\begin{align}
\Delta_{\pm}' &=\Psi \pm  \Delta_{d} \sin \theta, 
\end{align}
which is equivalent to that in the previous papers.\cite{NagaiCe,HayashiCe} 

Next, we assume the degenerated Fermi surface ($M = N = 2$), which is appropriate when $|\Vec{g}_{\Vec{k}}| \ll 1$. 
With the use of the unitary matrix $A(\phi,\theta)$ defined as 
\begin{align}
A(\phi,\theta) &\equiv \left(\begin{array}{cc}
e^{i \frac{\phi}{2}} & 0 \\
0 & -e^{-i \frac{\phi}{2}}
\end{array}\right), \label{eq:aphi}
\end{align}
the effective gap function can be rewritten as 
\begin{align}
\Delta_{\rm eff}(\phi,\theta)' &= \left(\begin{array}{cc}
\Psi + \Delta_{d} \sin \theta & 0 \\
0 & \Psi - \Delta_{d} \sin \theta
\end{array}\right). \label{eq:effectiveDelta}
\end{align}
In terms of the multi-band quasiclassical theory, we clarify that the decoupled equations (\ref{eq:ce}) are valid with the arbitrary strength spin-orbit coupling. 

Finally, we consider a specular reflection at a surface perpendicular to $x-y$ plane. 
We consider that the quasiparticles before and after a scattering have momentum $\Vec{k}_{1} = (\phi_{1},\theta)$,  $\Vec{k}_{2} = (\phi_{2},\theta)$, respectively.
By assuming the degenerated Fermi surface ($M = N = K = 2$), 
the unitary matrix with $A(\phi,\theta)$ in Eq.~(\ref{eq:aphi}) becomes 
\begin{align}
\tilde{U}_{\Vec{k}}^{M} &=\frac{1}{\sqrt{2}} \left(\begin{array}{cc}
-i e^{- i \phi/2} & -e^{- i \phi/2}  \\
e^{i \phi/2}  & i e^{ i \phi/2} 
\end{array}\right),
\end{align}
whose effective order parameter is given in Eq.~(\ref{eq:effectiveDelta}).
The transfer matrix $\tilde{V}_{\Vec{k}_{1}}^{\Vec{k}_{2}} \equiv \tilde{U}_{\Vec{k}_{2}}^{M \dagger}\tilde{U}_{\Vec{k}_{}}^{M}$ is expressed as
\begin{align}
\tilde{V}_{\Vec{k}_{1}}^{\Vec{k}_{2}} &= 
\left(\begin{array}{cc}
\cos \Delta \phi & -\sin \Delta \phi \\
\sin \Delta \phi & \cos \Delta \phi
\end{array}\right),
\end{align}
with $\Delta \phi \equiv(\phi_{1}-\phi_{2})/2$.
This transfer matrix suggests that both intra- and inter-band scatterings occur at a specular surface. 
The surface bound states and spin currents discussed in the previous paper\cite{VV} 
can be explained in terms of this band-active surface.

\subsection{Three-orbital model: Sr$_{2}$RuO$_{4}$}
\begin{figure}[t]
\begin{center}
\resizebox{0.6 \columnwidth}{!}{\includegraphics{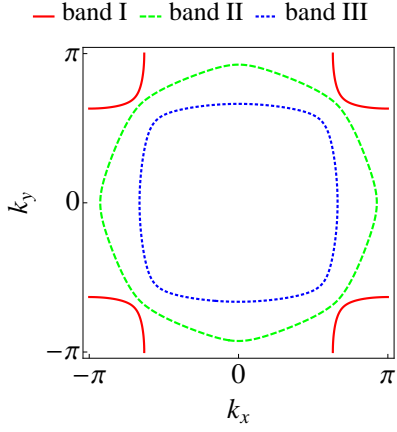}}
\end{center}
\caption{\label{fig:fermi}(Color online) Fermi surfaces for the three-band superconductor Sr$_{2}$RuO$_{4}$. 
}
\end{figure}

Let us apply our theory to a multi-band superconductor.
In this section, we consider Sr$_{2}$RuO$_{4}$ as the three-band superconductor. 
The many tight-binding models for Sr$_{2}$RuO$_{4}$ have been proposed by several authors\cite{Zabolotnyy, Kee,Ng,PuetterPRB,WCLee}. 
According to Ref.~\onlinecite{Zabolotnyy}, the effective tight-binding Hamiltonian is expressed by the three-orbital model characterized 
by $d^{yz}$-, $d^{xz}$-, and $d^{xy}$- orbitals ($N = 3$) expressed as 
\begin{align}
\check{H}(\Vec{k}) &= \left(\begin{array}{ccc}
\epsilon_{\Vec{k}}^{yz} - \mu & \epsilon_{\Vec{k}}^{\rm off} + i \lambda & - \lambda \\
\epsilon_{\Vec{k}}^{\rm off}- i \lambda & \epsilon_{\Vec{k}}^{xz} - \mu  & i \lambda \\
-\lambda & - i \lambda & \epsilon_{\Vec{k}}^{xy} - \mu 
\end{array}\right),
\end{align}
where
\begin{align}
\epsilon^{yz}_{\Vec{k}} &= -2 t_{2} \cos(k_{x}) - 2 t_{1} \cos (k_{y}), \nonumber \\
\epsilon^{xz}_{\Vec{k}} &= -2 t_{1} \cos(k_{x}) - 2 t_{2} \cos (k_{y}), \nonumber \\
\epsilon^{xy}_{\Vec{k}} &= -2 t_{3} (\cos(k_{x})+\cos(k_{y})) - 4 t_{4} \cos (k_{x}) \cos(k_{y}) \nonumber \\
&- 2 t_{5} (\cos (2 k_{x}) + \cos(2 k_{y})), \nonumber \\
\epsilon_{\Vec{k}}^{\rm off} &= -4 t_{6} \sin (k_{x}) \sin (k_{y}), 
\end{align}
with $\lambda = 0.032$, $t_{1} = 0.145$, $t_{2} = 0.016$, $t_{3} = 0.081$, $t_{4} = 0.039$, $t_{5} = 0.005$, $t_{6} = 0$, and 
$\mu = 0.122$. 
Here, we adopt the material parameters in Ref.~\onlinecite{Zabolotnyy}, which can successfully describe the three Fermi surfaces for 
Sr$_{2}$RuO$_{4}$ as shown in Fig.\ref{fig:fermi}. 
We call each band as band I, band II, and band III in ascending order of the eigenvalues.  

\begin{figure}[t]
\begin{center}
\resizebox{0.9 \columnwidth}{!}{\includegraphics{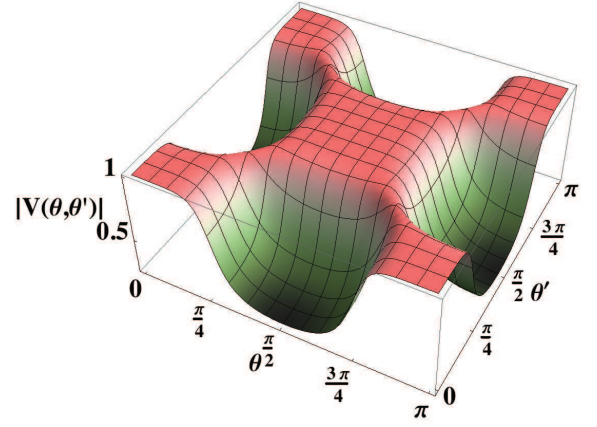}}
%\resizebox{0.7 \columnwidth}{!}{\includegraphics{Fig2r.eps}}
\end{center}
\caption{\label{fig:aniso}(Color online) The in-plane anisotropy of the effective potential $|V(\theta,\theta')|$ at the most inner Fermi surface in the three-band superconductor Sr$_{2}$RuO$_{4}$. 
}
\end{figure}

We consider the non-magnetic impurities to discuss the non-local anisotropic effective potential.  
We introduce the in-plane anisotropy of the effective potential defined as 
\begin{align}
V(\theta,\theta') \equiv \Vec{u}_{\Vec{k}_{\rm F}(\theta)}^{\dagger} \Vec{u}_{\Vec{k}_{\rm F}(\theta')},
\end{align}
Here, $\Vec{k}_{\rm F}(\theta)$ denotes the position of the most inner Fermi surface (i.e. band III) in momentum space ($\Vec{k}_{\rm F}(\theta) =k_{\rm F}(\theta)(\cos \theta,\sin \theta)$).
As shown in Fig.~\ref{fig:aniso}, the right-angled scatterings suppress in the most inner Fermi surface. 
This suppression originates from the fact that the eigenvector associated with $\Vec{k}_{\rm F}(\theta=0)$ mainly consists of $d_{xz}$ orbital and the eigenvector associated with $\Vec{k}_{\rm F}(\theta'=0)$ mainly consists of $d_{yz}$ orbital. 

\subsection{Heavy fermion CeCoIn$_{5}$/YbCoIn$_{5}$ superlattice: The perturbative approach}
In this section, we consider the system with both the spin-orbit coupling and the Zeeman interaction. 
The locally noncentrosymmetric systems are realized in the heavy fermion CeCoIn$_{5}$/YbCoIn$_{5}$ superlattice\cite{Mizukami}. 
In these systems, the layer-dependent spin-orbit coupling induces the exotic superconducting states.  
In the $N$-layer spin-singlet $s$-wave superconductor, 
the multi-band Eilenberger equations with $4 N \times 4 N$ matrix quasiclassical Green's function are 
written as 
%\begin{widetext}
\begin{align}
 i \Vec{v}_{\rm F}\cdot \Vec{\nabla}
  \check{g}   +
  \left[
 z  \check{\tau}_{z}
    -
   \check{\Delta}
 -
\check{K} (\Vec{k}_{\rm F}) 
,
 \bar{g}
   \right]_{-}
&= 0, 
\end{align}
with 
\begin{align}
\check{K} (\Vec{k}_{\rm F}) &\equiv 
\left( t_{\perp} \hat{H}_{\rm inter}+\mu_{\rm B} h \hat{H}_{Z}\right) \check{\tau}_{0} \nonumber \\
&+  \left(\begin{array}{cc}
\hat{H}_{\rm SO}(\Vec{k}_{\rm F}) & 0 \\
0 & \hat{H}_{\rm SO}^{\ast}(-\Vec{k}_{\rm F}) 
\end{array}\right), \\
\check{\tau}_{z} &= \left(\begin{array}{cc}
\sigma_{0} \otimes I_{N \times N} & 0 \\
0 & -\sigma_{0} \otimes I_{N \times N} 
\end{array}\right) \\
 \check{\Delta} &= \left(\begin{array}{cc}
 0 & \hat{\Delta} \\
- \hat{\Delta}^{\dagger} & 0
 \end{array}\right), \\
 \hat{H}_{\rm inter} &= \sigma_{0} \otimes T_{\perp}, \\
 \hat{H}_{Z} &= - \sigma_{z} \otimes I_{N \times N}, \\
 \hat{H}_{\rm SO}(\Vec{k}_{\rm F}) &= \Vec{g}(\Vec{k}_{\rm F}) \cdot \Vec{\sigma} \otimes S_{d}.
\end{align}
Here, $\sigma_{z}$ is the $2 \times 2$ Pauli matrix, $I_{N \times N}$ is the unit matrix, $T_{\perp}$ is the hopping matrix between layers, and 
$S_{d} = {\rm diag} (\alpha_{i},\cdots,\alpha_{N})$ denotes the layer-dependent spin-orbit interaction. 
In the above equations, the hopping, the Zeeman, and the spin-orbit coupling terms are regarded as the perturbations with 
$2N$-degenerated Fermi surfaces. 
With the use of this perturbation theory, one can treat the inhomogeneous system with vortices\cite{HigashiLT}.  
%\end{widetext}

\subsection{Topological superconductors with the strong spin-orbit coupling: Cu$_{x}$Bi$_{2}$Se$_{3}$}
We discuss the boundary condition in the three-dimensional topological superconductor with the strong spin-orbit coupling in this section. 
Cu$_{x}$Bi$_{2}$Se$_{3}$ is the one of the candidates of the topological superconductors where the topologically protected 
Majorana bound states form at the boundary. 
We have proposed the quasiclassical framework on topological superconductors with strong spin-orbit coupling\cite{NagaiTopo}.
In the previous paper\cite{NagaiTopo}, we have obtained the linearized BdG equations called Andreev equations 
by decomposing the slow varying component from the total quasi-particle wave function. 
Applying this quasiclassical treatment, the original massive Dirac BdG Hamiltonian 
derived from the tight-binding model represented by $8 \times 8$ matrix 
is reduced to $4 \times 4$ one. 
The resultant Andreev equations become equivalent to those of spin-singlet or triplet superconductors without the spin-orbit coupling. 
In this section, we show that the same result is obtained by the Green's function techniques. 

The normal-states effective Hamiltonian for  Cu$_{x}$Bi$_{2}$Se$_{3}$ is expressed as 
\begin{align}
\hat{H}(\Vec{k}) &= 
\left(\begin{array}{cc}
(M(\Vec{k}) - \mu ) \sigma_{0} & \Vec{k}\cdot \Vec{\sigma} \\
\Vec{k}\cdot \Vec{\sigma} & (-M(\Vec{k}) - \mu) \sigma_{0} \end{array}\right). \label{eq:topo}
\end{align}
Here $\sigma_{i}$ denotes the Pauli matrix in the spin space. 
In the quasiclassical theory, $\hat{H}(\Vec{k})$ is regarded as $\hat{H}^{\rm N 1}(\Vec{k})$ in Eq.~(\ref{eq:eigen}). 
The eigenvalues of the $4 \times 4$ matrix are degenerated expressed as 
\begin{align}
\epsilon_{i}(\Vec{k}) &= \pm E_{0}(\Vec{k}) - \mu, 
\end{align}
where $E_{0}(\Vec{k}) \equiv \sqrt{M(\Vec{k})^{2} + |\Vec{k}|^{2}}$. 
In the case of $\mu > 0$, the  eigenvectors $u_{\Vec{k}}^{1}$ and $u_{\Vec{k}}^{2}$ ($M=2$),  and eigenvalue $\xi_{\Vec{k}}$
in Eq.~(\ref{eq:eigen}) are given as 
\begin{align}
\xi_{\Vec{k}} &= E_{0}(\Vec{k}) - \mu \\
u_{\Vec{k}}^{i} &= c \left(\begin{array}{c}
\chi_{i} \\
\frac{\Vec{k} \cdot \Vec{\sigma}}
{E_{0}(\Vec{k})+ M(\Vec{k})}
\chi_{i}
\end{array}\right),
\end{align}
where $\chi^{\rm T}_{1} \equiv (1,0)$, $\chi^{\rm T}_{2} \equiv (0,1)$, $c \equiv \sqrt{(E_{0}+M(\Vec{k}))/2 E_{0}}$. 
We consider the $4 \times 4$ odd-parity fully-gapped gap function (so-called $A_{1u}$ state) defined as 
\begin{align}
\hat{\Delta} &\equiv \Delta_{0} \left(\begin{array}{cc}
0 & i \sigma_{y} \\
i \sigma_{y} & 0
\end{array}\right). \label{eq:gappseudo}
\end{align}
By substituting this gap function into Eq.~(\ref{eq:gapeff}), we obtain the $2 \times 2$ effective gap function 
written as 
\begin{align}
\Delta_{\rm eff}(\Vec{k}) &=  \frac{\Delta_{0}}{E_{0}(\Vec{k})}  \Vec{k} \cdot \Vec{\sigma} (i \sigma_{y}). 
\end{align}
This gap function is completely equivalent to that in the previous paper \cite{NagaiTopo} in terms of the Dirac BdG Hamiltonian. 

Let us consider the boundary condition with a specular surface at $z = 0$. 
We adopt the boundary condition that all components of the wave-function becomes zero at $z =0$, which is different from that discussed in Ref.~\onlinecite{NagaiQuasi}. 
We consider $M(\Vec{k}) = M_{0}(\Vec{k}_{\parallel}) + M_{1} k_{z}^{2}$.
By using Eq.~(\ref{eq:boundnh}), we find that the wave numbers become 
\begin{align}
(k_{z \pm})^{2} = \frac{-1}{2 M_{1}^{2}} \left[ 1 + 2 M_{0} M_{1} \pm \xi_{\Vec{k}_{\parallel}}
 \right],
 \end{align}
 with $\xi_{\Vec{k}_{\parallel}} \equiv \sqrt{
(1 + 2 M_{0} M_{1})^{2} + 4 M_{0} (\mu^{2} - M_{0}^{2} - |\Vec{k}_{\parallel}|^{2})}$. 
When the condition $\mu >\sqrt{ M_{0}^{2} - |\Vec{k}_{\parallel}|^{2}}$ is satisfied, 
there are two real wave numbers and two imaginary wave number, expressed as 
\begin{align}
k_{z -}^{1} &= k_{-} %\sqrt{\frac{-1}{2 M_{1}^{2}} \left[ 1 + 2 M_{0} M_{1} -\xi_{\Vec{k}_{\perp}}
\\
k_{z -}^{2} &= -k_{-}%\sqrt{\frac{-1}{2 M_{1}^{2}} \left[ 1 + 2 M_{0} M_{1} - \xi_{\Vec{k}_{\perp}}
, \\
 k_{z +}^{1} &= i \eta_{+}%\sqrt{\frac{1}{2 M_{1}^{2}} \left[ 1 + 2 M_{0} M_{1} + \xi_{\Vec{k}_{\perp}}
, \\
 k_{z +}^{2} &= -i \eta_{+}%\sqrt{\frac{1}{2 M_{1}^{2}} \left[ 1 + 2 M_{0} M_{1} + \xi_{\Vec{k}_{\perp}}
,
\end{align}
with $k_{-} \equiv \sqrt{-\left[ 1 + 2 M_{0} M_{1} -\xi_{\Vec{k}_{\parallel}}  \right]/(2 M_{1}^{2})}$ and 
$\eta_{+} \equiv \sqrt{\left[ 1 + 2 M_{0} M_{1} + \xi_{\Vec{k}_{\parallel}}  \right]/(2 M_{1}^{2})}$. 
By assuming that the material is filled in the region $z > 0$, the coefficient of the wavefunction with $k_{z +}^{2}$ is zero. 
Thus, we obtain $K = 3$. 
The boundary condition (\ref{eq:nbound}) is expressed as 
\begin{align}
&\left(\begin{array}{ccc}\Vec{u}^{1}_{(\Vec{k}_{\parallel},k_{-})} &  \Vec{u}^{1}_{(\Vec{k}_{\parallel},-k_{-})} & \Vec{u}^{1}_{(\Vec{k}_{\parallel},i \eta_{+})} 
\end{array}\right) \Vec{c}^{1} \nonumber \\
&+
\left(\begin{array}{ccc}\Vec{u}^{2}_{(\Vec{k}_{\parallel},k_{-})} &  \Vec{u}^{2}_{(\Vec{k}_{\parallel},-k_{-})} & \Vec{u}^{2}_{(\Vec{k}_{\parallel},i \eta_{+})} 
\end{array}\right) \Vec{c}^{2}
=0.
\end{align}
When we consider $\Vec{k}_{\parallel} = 0$, the coefficients are given as 
\begin{align}
 \Vec{c}^{1} &=  \Vec{c}^{2} = \left(\begin{array}{c}
 \frac{1}{2} \left(1 + \frac{\mu + M(k_{-})}{\mu + M(i \eta_{+})} \frac{i \eta_{-}}{k_{-}} \right) \\
 \frac{1}{2} \left(1 - \frac{\mu + M(k_{-})}{\mu + M(i \eta_{+})} \frac{i
  \eta_{-}}{k_{-}} \right) \\
-1
 \end{array}\right) c.
\end{align}
Thus, the boundary condition for the quasiclassical wave function is obtained as 
\begin{align}
\Vec{f}^{-k_{-}}(z=0) &= c_{1}  \Vec{f}^{k_{-}}(z=0), \\
\Vec{g}^{-k_{-}}(z=0) &= c_{1}^{\ast} \Vec{g}^{k_{-}}(z=0), 
\end{align}
where
\begin{align}
c_{1} &=
\frac{
k_{-} (\mu + M(i \eta_{+})) - i \eta_{+} (\mu + M(k_{-})) 
}
{
k_{-} (\mu + M(i \eta_{+})) + i \eta_{+} (\mu + M(k_{-})) 
}.
\end{align}
In the non-relativistic limit ($|k_{-}| \ll M(k_{-})$), 
the boundary condition becomes
\begin{align}
\Vec{f}^{-k_{-}}(z=0) &=  -\Vec{f}^{k_{-}}(z=0), \\
\Vec{g}^{-k_{-}}(z=0) &=  -\Vec{g}^{k_{-}}(z=0), 
\end{align}
which is equivalent to that in a single band quasiclassical framework. 
This result is consistent with the fact that this superconductor becomes a $p$-wave superconductor in the non-relativistic limit\cite{Nagai3D}.

\subsection{Robust $p$-wave superconductivity on a surface of topological insulator}
\label{sec:rp}
In this section, we discuss the impurity effect in the $s$-wave gap superconductivity on a surface of topological insulator. 
Let us show that the proximity-induced superconductivity on a surface of topological insulator is robust against nonmagnetic impurities\cite{Ito:2011ct}. 
We consider that an $s$-wave superconductor is deposited on the surface of the topological insulator\cite{Fu:2008gu}.
The effective two-dimensional Hamiltonian on the surface is described as 
\begin{align}
\hat{H}(\Vec{k}) &= \left(\begin{array}{cc}
h_{0}(\Vec{k})& \Delta i \sigma_{y} \\
\Delta^{\ast}(- i \sigma_{y}) & 
-h_{0}^{\ast}(-\Vec{k})
\end{array}\right),
\end{align}
with 
\begin{align}
h_{0}(\Vec{k}) &\equiv v \Vec{k} \cdot \Vec{\sigma} -\mu \sigma_{0}. 
\end{align}
The eigenvalues of the $2 \times 2$ normal state Hamiltonian $h_{0}(\Vec{k})$ are given by 
\begin{align}
\epsilon_{\pm}(\Vec{k}) &= - \mu \pm v |\Vec{k}|.
\end{align}
The eigenvectors associated with $\epsilon_{\pm}(\Vec{k})$ are expressed as 
\begin{align}
u_{\Vec{k}}^{+} &= 
\frac{1}{\sqrt{2}}
 \left(\begin{array}{c}
e^{- i \phi} \\
1
\end{array}\right), \\
u_{\Vec{k}}^{-} &= 
\frac{1}{\sqrt{2}}
 \left(\begin{array}{c}
1 \\
-e^{ i \phi}
\end{array}\right),
\end{align}
where $\Vec{k} = (k_{x},k_{y})= k (\cos \phi, \sin \phi)$. 
In the case of $\mu > 0$, the $1 \times 1$ effective gap is given as 
\begin{align}
\Delta_{\rm eff}(\Vec{k}) &= \Delta e^{i \phi}. \label{eq:peff}
\end{align}
Thus, the effective $p$-wave superconductivity appears on a surface of the topological insulator.

Let us consider a nonmagnetic impurity effect in this proximity-induced superconductor. 
The non-perturbative quasiclassical Green function in a homogeneous system is given as 
\begin{align}
\bar{g}(\Vec{k}_{\rm F}, i \omega_{n}) &= \frac{- \pi}{\sqrt{\omega_{n}^{2} + |\Delta|^{2}}}
\left(\begin{array}{cc}
i \omega_{n} & \Delta e^{i \phi} \\
- \Delta e^{-i \phi} & - i \omega_{n}
\end{array}\right).
\end{align}
The effective potential $\bar{V}(\Vec{k}_{\rm F},\Vec{k}_{\rm F}')$ in Eq.~(\ref{eq:vmulti}) is expressed as 
\begin{align}
\bar{V}(\Vec{k}_{\rm F},\Vec{k}_{\rm F}') &= V  \left(\begin{array}{cc}
e^{i \delta \phi/2} \cos \left( \frac{\delta \phi}{2}  \right) & 0 \\
0 & e^{-i \delta \phi /2} \cos \left( \frac{\delta \phi}{2}  \right)
\end{array}\right),
\end{align}
with $\delta \phi \equiv \phi - \phi'$. 
Here, $V$ is the amplitude of the potentials. 
The second order of the impurity self-energy in Eq.~(\ref{eq:sigma}) becomes 
\begin{align}
\bar{\Sigma}(\Vec{k}_{\rm F}, i \omega_{n})^{(2)}\bar{\sigma}_{z} &= n_{\rm imp} V^{2}\bar{g}(\Vec{k}_{\rm F}, i \omega_{n})
 \int_{0}^{2 \pi} \frac{d\phi' \cos \left(\frac{\phi - \phi'}{2} \right)}{2 \pi}. 
\end{align}
Since this self-energy satisfies $[\bar{\Sigma}(\Vec{k}_{\rm F}, i \omega_{n})^{(2)}\bar{\sigma}_{z} ,\bar{g}(\Vec{k}_{\rm F}, i \omega_{n})]_{-}=0$, 
the quasiclassical Eilenberger equations with the self-energy are completely equivalent to those without the self-energy. 
Therefore, this proximity-induced superconductivity on the surface of a topological insulator is robust against nonmagnetic impurities. 

Finally, we point out that the previous decoupled quasiclassical framework can not reproduce the robustness against nonmagnetic impurities proposed in Ref.~\onlinecite{Ito:2011ct}. 
With the use of the band basis, the effective gap is given in Eq.~(\ref{eq:peff}) at the Fermi energy. 
Since this effective gap means $p$-wave superconductivity, the superconductivity should be
 fragile against nonmagnetic impurities in the previous decoupled quasiclassical framework.

\subsection{Surface quasiclassical theory: the partial quasiclassical approximation for topological insulators}
Let us discuss the ``partial'' quasiclassical approximation with considering the topological insulators. 
The superconductivity in surface states on topological insulators has been attracted much attention because of the stage of the Majorana Fermion 
and a quantum computing.
With the use of the proximity effects from the superconductor on the topological insulator, 
the two-dimensional massless Dirac quasiparticles due to the surface bound states on the topological insulator  
form the superconducting Cooper pairs. 
Therefore, it is important to construct the two-dimensional effective Eilenberger equations originating from the 
normal-state surface bound states. 

Let us consider the surface at $z = 0$. 
By introducing the coordinate $\Vec{r} = (x,y,z) \equiv (\Vec{r}_{\perp},z)$, 
we can define the partial Wigner representation expressed as 
\begin{align}
& \check{A}_{z_{1}z_{2}}(\Vec{R}_{\perp},\Vec{k}_{\perp}) \equiv \nonumber \\
& \int d\Vec{\bar{r}}_{\perp} e^{-i \Vec{k}_{\perp} \cdot \Vec{\bar{r}}_{\perp}} \check{A}_{z_{1}z_{2}}\left( \Vec{R}_{\perp} + \frac{\Vec{\bar{r}}_{\perp} }{2},\Vec{R}_{\perp} - \frac{\Vec{\bar{r}}_{\perp}}{2} \right).
\end{align}
Here, $\Vec{R}_{\perp} = (\Vec{r}_{1 \perp}+\Vec{r}_{2 \perp})/2$ and $\Vec{\bar{r}}_{\perp}= \Vec{r}_{1 \perp} - \Vec{r}_{2 \perp}$ are the center-of-mass coordinate and the relative 
coordinate on the two-dimensional plane parallel to the surface, respectively. 
The projection operator is determined as 
\begin{align}
\check{P}_{\Vec{k}_{\perp} z_{1}z_{2}} &\equiv U_{\Vec{k}_{\perp}}^{M}(z_{1}) U_{\Vec{k}_{\perp}}^{M \dagger}(z_{2}),
\end{align}
where 
\begin{align}
U_{\Vec{k}_{\perp}}^{M}(z)&\equiv 
\left(\begin{array}{cc}
\tilde{U}_{\Vec{k}_{\perp}}^{M}(z) & 0 \\
0 &  \tilde{U}_{-\Vec{k}_{\perp}}^{M \ast}(z) 
\end{array}\right), 
\end{align}
with 
the matrix $U_{\Vec{k}_{\perp}}^{M}(z) = (u_{\Vec{k}_{\perp}}^{1}(z),\cdots,u_{\Vec{k}_{\perp}}^{M}(z))$. 
Here, the vector $u_{\Vec{k}}^{i}(z)$ is the eigenvector expressed as 
\begin{align}
\hat{H}^{\rm N 1}(\Vec{k}_{\perp},z) u_{\Vec{k}_{\perp}}^{i}(z) &= \xi_{\Vec{k}_{\perp}} u_{\Vec{k}_{\perp}}^{i}(z).
\end{align}
We note that the Hamiltonian $\hat{H}^{\rm N 1}(\Vec{k},z)$ includes information about a presence of a surface. 
The projected effective gap function is given by 
\begin{align}
\bar{\Delta}(\Vec{R}_{\perp},\Vec{k}_{\perp}) &\equiv \int dz_{1} dz_{2} U_{\Vec{k}_{\perp}}^{M \dagger}(z_{1}) \check{\Delta}_{z_{1} z_{2}}(\Vec{R}_{\perp},
\Vec{k}_{\perp})U_{-\Vec{k}_{\perp}}^{M \ast}(z_{2}), \label{eq:zgap}
\end{align}
since the projection includes the $z$-integration written as  
\begin{align}
\check{A}_{z_{1} z_{2}}' &=
\int dz_{3} \check{P}_{\Vec{k} z_{1}z_{3}} \check{A}_{z_{3} z_{2}}. 
\end{align}
Let us consider the three dimensional topological superconductor as an example. 
The eigenvector in Eq.~(\ref{eq:topo}) with the boundary condition $u_{\Vec{k}_{\perp}}^{i}(z=0) = 0$ is expressed 
as\cite{SCES} 
\begin{align}
u_{\Vec{k}_{\perp}}^{i}(z) &= \frac{e^{\frac{z}{2 M_{1}}} \sinh(K z)}{2 \sqrt{A}} 
\left(\begin{array}{c}
e^{- i \phi} \\
i \\
i e^{- i \phi} \\
1
\end{array}\right), \label{eq:zu}
\end{align}
where $\xi_{\Vec{k}_{\perp}} = \sqrt{k_{x}^{2}+k_{y}^{2}} - \mu$, $\Vec{k}_{\perp} = (k_{x},k_{y}) =\sqrt{k_{x}^{2}+k_{y}^{2}} (\cos \phi,\sin \phi)$, $M(\Vec{k}) = M_{0}(\Vec{k}_{\perp}) + M_{1} k_{z}^{2} $, $K = (\sqrt{1+4 M_{0} M_{1}})/(2 M_{1})$, 
and $A = \int_{0}^{\infty} dz exp(z/M_{1}) |\sinh(Kz)|^{2}$.
Here, we assume $\mu > 0$ and obtain the above solution with the use of the perturbation with respect to $\Vec{k}_{\perp}$.  
By substituting the eigenvector in Eq.~(\ref{eq:zu}) and the odd-parity fully-gapped gap function in Eq.~(\ref{eq:gappseudo}) into 
Eq.~(\ref{eq:zgap}), 
we obtain 
\begin{align}
\bar{\Delta}(\Vec{R}_{\perp},\Vec{k}_{\perp}) &= \bar{0}.
\end{align}
Therefore, the proximity-induced odd-parity fully-gapped gap function does not open the spectral gap as shown in Ref.~\onlinecite{SCES}.  

With the use of the above method, we directly show that the proximity-induced $s$-wave superconductivity on the topological insulator can be regarded as 
a chiral $p$-wave superconductivity, as shown in Sec.~\ref{sec:rp}. 
By substituting the eigenvector in Eq.~(\ref{eq:zu}) and the even-parity fully-gapped spin-singlet intra-orbital gap function expressed as 
\begin{align}
\hat{\Delta} &\equiv \Delta_{0} \left(\begin{array}{cc}
 i \sigma_{y} & 0\\
0 &  i \sigma_{y}
\end{array}\right), 
\end{align}
into 
Eq.~(\ref{eq:zgap}), 
we obtain 
\begin{align}
\bar{\Delta}(\Vec{R}_{\perp},\Vec{k}_{\perp}) &= \Delta_{\perp} e^{i \phi},
\end{align}
which is equivalent to the chiral $p$-wave superconductivity in Eq.~(\ref{eq:peff}).

\subsection{Others}
Finally, we discuss the advantage of the multi-band quasiclassical theory. 
The computational cost drastically decreases with the use of our theory in multi-band systems.
Thus, we can  treat the inhomogeneous systems such as those with vortices and surfaces easily, in order to 
discuss the magnetic field dependence of the multi-band superconductors. 
Since we do not use any assumptions about the electronic structures in normal states, the superconducting system with the arbitrary 
tight-binding Hamiltonian derived by the first-principle calculation can be mapped onto the effective low-energy system. 
In the theoretical point of view, one might develop the general theory for impurity effects in multi-band superconductors, 
since the multi band effects are explicitly included as the non-local and anisotropic potentials. 
It should be noted that one can understand what is neglected in the quasiclassical theory in multi-band superconductors. 
One might know the difference between the single-band and multi-band superconductors through the study with our multi-band Eilenberger theory.

\section{Summary}
\label{sec:sum}

In summary, we proposed the unified quasiclassical multi-band Eilenberger equations in order to map the multi-band systems onto the effective 
systems in the reduced space. 
We derived both the Andreev and Eilenberger equations with an arbitrary boundary condition. 
We showed that the resultant multi-band Eilenberger equations are similar to the single-band ones, except for some corrections to describe multi-band effects. 
The orbital characters on the Fermi surfaces in normal states are included in our theory. 
Our theory could describe the past studies with the use of the quasiclassical Eilenberger theory. 
Since we do not use any assumptions about the electronic structures in normal states, the superconducting system with the arbitrary 
tight-binding Hamiltonian derived by the first-principle calculation can be mapped onto the effective low-energy system. 
The potentials, the order-parameters, and self-energies in multi-band systems were mapped onto the non-local ones in the reduced space as shown in Eqs.~(\ref{eq:greenu})-(\ref{eq:sigmau}). 
We showed that the self-energy with the $T$-matrix approximation of the non-magnetic impurities becomes non-local and anisotropic. 
We pointed out that this non-locality is similar to the pseudo potential in the first-principles calculations. 
The multi-band effects can be understood by the non-locality and the anisotropy in the mapped systems.

\section*{Acknowledgment}
The authors would like to acknowledge Masahiko Machida, Susumu Yamada, Yusuke Kato, Yukihiro Ota, Kaori Tanaka and Nobuhiko Hayashi for 
helpful discussions and comments. 
The calculations have been performed using the supercomputing system PRIMERGY BX900 at the Japan Atomic Energy Agency.
This study was supported by JSPS KAKENHI Grant Number 26800197.

\appendix

\section{Derivation of the Andreev equations}
\label{app:and}
We derive the multi-band Andreev-type equations as follows. 
We substitute Eq.(\ref{eq:wave}) into the BdG equations (\ref{eq:BdGeq}). 
\begin{widetext}
The diagonal blocks in the BdG equations become 
%%%
\begin{align}
 \hat{H}^{\rm N}(\Vec{r}_{1},- i \Vec{\nabla}_{1}) \left[ e^{i \Vec{k}_{\rm F} \Vec{r}_{1}} f_{l}^{\Vec{k}_{\rm F}}(\Vec{r}_{1})  \Vec{u}_{l}^{\rm N}(\Vec{k}_{\rm F})
\right] \sim  e^{i \Vec{k}_{\rm F} \Vec{r}_{1}}  \left[ 
 \hat{H}^{\rm N 0}(\Vec{r}_{1})  f_{l}^{\Vec{k}_{\rm F}}(\Vec{r}_{1}) 
%\right. \nonumber \\
%& \left.
 - i  
\frac{\partial \hat{H}^{\rm N 1}(\Vec{k})}{\partial \Vec{k}} \Big|_{\Vec{k} = \Vec{k}_{\rm F}}  \cdot \Vec{\nabla} f_{l}^{\Vec{k}_{\rm F}}(\Vec{r}_{1}) 
 \right] \Vec{u}_{l}^{\rm N}(\Vec{k}_{\rm F}) .
\end{align}
%%%
The off-diagonal blocks are converted as 
%%%
\begin{align}
\int d \Vec{r}_{2} \hat{\Delta}(\Vec{r}_{1},\Vec{r}_{2}) \left[ e^{i \Vec{k}_{\rm F} \Vec{r}_{2}} g_{l}^{\Vec{k}_{\rm F}}(\Vec{r}_{2})  \Vec{v}_{l}^{\rm N}(\Vec{k}_{\rm F})
\right] &\sim e^{i \Vec{k}_{\rm F} \Vec{r}_{1}} \hat{\Delta}(\Vec{r}_{1},\Vec{k}_{\rm F}) g_{l}^{\Vec{k}_{\rm F}}(\Vec{r}_{1})  \Vec{v}_{l}^{\rm N}(\Vec{k}_{\rm F}), 
\end{align}
%%%
where $\hat{\Delta}(\Vec{r}_{1},\Vec{k}_{\rm F})$ is the order parameter matrix in the Wigner representation. 
\end{widetext}
With the use of the relation 
%%%%%
\begin{align}
 \tilde{U}_{\Vec{k}}^{M \dagger} \frac{\partial \hat{H}^{\rm N 1}(\Vec{k})}{\partial \Vec{k}} 
 \tilde{U}_{\Vec{k}}^{M} \Big|_{\Vec{k} = \Vec{k}_{\rm F}}  &= 
  \frac{\partial \xi(\Vec{k})}{\partial \Vec{k}} \Big|_{\Vec{k} = \Vec{k}_{\rm F}} 1_{M \times M} \equiv \Vec{v}_{\rm F} 1_{M \times M},  
\end{align}
we obtain the multi-band Andreev-type equations (\ref{eq:and}).
%%%%%

\section{Expansion of the $\star$-products}
\label{sec:expansion}
We expand the $\star$-products for $\check{\Delta}(\Vec{R},\Vec{k})$ and $\check{\Sigma}(\Vec{R},\Vec{k})$ 
written as 
\begin{align}
& \left[ 
\check{\Delta}(\Vec{R},\Vec{k}) +\check{\Sigma}(\Vec{R},\Vec{k}) \right] \star  \check{G}(\Vec{R},\Vec{k},i \omega_{n})  \nonumber  \\
&\sim  \left[ 
\check{\Delta}(\Vec{R},\Vec{k}) +\check{\Sigma}(\Vec{R},\Vec{k}) \right] \check{G}(\Vec{R},\Vec{k},i \omega_{n}). 
\end{align}
In the quasiclassical theory of superconductivity, the $\star$-products for $\check{H}^{\rm N0}(\Vec{R})$ and $\check{H}^{\rm N1}(\Vec{k})$ 
are expanded in the 1st order written as 
%%%%%%%
\begin{align}
\check{H}^{\rm N 0}(\Vec{R}) \star  \check{G}(\Vec{R},\Vec{k},i \omega_{n}) &\sim 
\check{H}^{\rm N 0}(\Vec{R})\check{G}(\Vec{R},\Vec{k},i \omega_{n})  \nonumber \\
&+\frac{i}{2 } 
 \Vec{\nabla}_{\Vec{R}}  \check{H}^{\rm N 0}(\Vec{R}) \cdot \Vec{\nabla}_{\Vec{k}} \check{G}(\Vec{R},\Vec{k},i \omega_{n}), \\
 \check{H}^{\rm N 1}(\Vec{k}) \star  \check{G}(\Vec{R},\Vec{k},i \omega_{n}) &\sim 
\check{H}^{\rm N 1}(\Vec{k})\check{G}(\Vec{R},\Vec{k},i \omega_{n})  \nonumber \\
&-\frac{i}{2 } 
 \Vec{\nabla}_{\Vec{k}}  \check{H}^{\rm N 1}(\Vec{k}) \cdot \Vec{\nabla}_{\Vec{R}} \check{G}(\Vec{R},\Vec{k},i \omega_{n}). 
\end{align}
%%%%%%
%\section{Derivation of the boundary condition about a specular surface}

%%%
\section{Derivation of the multi-band quasiclassical Eilenberger equations}
\label{sec:quasieilen}
We derive the multi-band quasiclassical Eilenberger equations. 
\begin{widetext}
By multiplying the the both sides in Eq.~(\ref{eq:wig1}) by the matrices $U_{\Vec{k}}^{M \dagger}$ and $U_{\Vec{k}}^{M}$, 
we obtain
%%% 
\begin{align}
& \left(
i \omega_{n} - 
\bar{V}_{0}(\Vec{R},\Vec{k}) -\xi_{\Vec{k}} \bar{\sigma}_{z} 
  - \bar{\Delta}(\Vec{R},\Vec{k}) - \bar{\Sigma}(\Vec{R},\Vec{k},i \omega_{n})
 \right) 
 \bar{G}(\Vec{R},\Vec{k},i \omega_{n}) \nonumber \\
& - \frac{i}{2 } 
 \Vec{\nabla}_{\Vec{R}}  \bar{V}_{0}(\Vec{R},\Vec{k}) \cdot U_{\Vec{k}}^{M \dagger} \left[  \Vec{\nabla}_{\Vec{k}}  \check{G}(\Vec{R},\Vec{k},i \omega_{n}) \right] U_{\Vec{k}}^{M}
 + \frac{i}{2 } 
U_{\Vec{k}}^{ M \dagger} \left[ \Vec{\nabla}_{\Vec{k}}  \check{H}^{\rm N 1}(\Vec{k}) \right] U_{\Vec{k}}^{M} \cdot \Vec{\nabla}_{\Vec{R}} \bar{G}(\Vec{R},\Vec{k},i \omega_{n})
 = \bar{1}. \label{eq:wig2}
\end{align}
%\end{widetext}
%%%%%%%
The right-hand Gor'kov equation in the projected space are written as 
%\begin{widetext}
%%% 
\begin{align}
&  \bar{G}(\Vec{R},\Vec{k},i \omega_{n}) \left(
i \omega_{n} - 
\bar{V}_{0}(\Vec{R},\Vec{k}) -\xi_{\Vec{k}}\bar{\sigma}_{z}  
  - \bar{\Delta}(\Vec{R},\Vec{k}) - \bar{\Sigma}(\Vec{R},\Vec{k},i \omega_{n})
 \right) 
 \nonumber \\
& + \frac{i}{2 } U_{\Vec{k}}^{M \dagger} \left[  \Vec{\nabla}_{\Vec{k}}  \check{G}(\Vec{R},\Vec{k},i \omega_{n}) \right] U_{\Vec{k}}^{M} \cdot
 \Vec{\nabla}_{\Vec{R}}  \bar{V}_{0}(\Vec{R},\Vec{k}) 
 - \frac{i}{2 } \Vec{\nabla}_{\Vec{R}} \bar{G}(\Vec{R},\Vec{k},i \omega_{n}) \cdot 
U_{\Vec{k}}^{ M \dagger} \left[ \Vec{\nabla}_{\Vec{k}}  \check{H}^{\rm N 1}(\Vec{k}) \right] U_{\Vec{k}}^{M} 
 = \bar{1}. \label{eq:wig2right}
\end{align}
%%%%%%%
By subtracting the right-hand Gor'kov equation, the equation (\ref{eq:wig2}) become 
%%%
\begin{align}
&\left[ i \omega_{n} \bar{\sigma}_{z}  - \bar{V}_{0}(\Vec{R},\Vec{k})  \bar{\sigma}_{z}  - \bar{\Delta}(\Vec{R},\Vec{k}) \bar{\sigma}_{z} 
- \bar{\Sigma}(\Vec{R},\Vec{k}) \bar{\sigma}_{z} 
,  \bar{\sigma}_{z} \bar{G}(\Vec{R},\Vec{k};i \omega_{n})  \right]_{-} \nonumber \\
& -\frac{i}{2} \left[ \Vec{\nabla}_{\Vec{R}} \bar{V}_{0}(\Vec{R},\Vec{k})  \bar{\sigma}_{z}, \bar{\sigma}_{z} 
U_{\Vec{k}}^{M \dagger} \left[  \Vec{\nabla}_{\Vec{k}}  \check{G}(\Vec{R},\Vec{k},i \omega_{n}) \right] U_{\Vec{k}}^{M} 
\right]_{+} 
+\frac{i}{2} \left[U_{\Vec{k}}^{ M \dagger} \left[ \Vec{\nabla}_{\Vec{k}}  \check{H}^{\rm N 1}(\Vec{k}) \right] U_{\Vec{k}}^{M}\bar{\sigma}_{z} ,   \bar{\sigma}_{z}  \Vec{\nabla}_{\Vec{R}} \bar{G}(\Vec{R},\Vec{k};i \omega_{n})  \right]_{+} = 0, 
\end{align}
\end{widetext}
where $[A,B]_{\pm} \equiv AB \pm BA$. 

In the quasiclassical theory, the information on the Fermi surface is most important. 
By assuming that $\bar{V}_{0}(\Vec{R},\Vec{k})$, $\bar{\Delta}(\Vec{R},\Vec{k})$, $\bar{\Sigma}(\Vec{R},\Vec{k})$, and $U_{\Vec{k}}^{M}$ 
are slowly varying functions around the Fermi momentum $\Vec{k}_{\rm F}$, 
we obtain  
\begin{widetext}
%%%
\begin{align}
&\left[ i \omega_{n} \bar{\sigma}_{z}  - \bar{V}_{0}(\Vec{R},\Vec{k}_{\rm F})  \bar{\sigma}_{z}  - \bar{\Delta}(\Vec{R},\Vec{k}_{\rm F}) \bar{\sigma}_{z} 
- \bar{\Sigma}(\Vec{R},\Vec{k}_{\rm F}) \bar{\sigma}_{z} 
,  \bar{\sigma}_{z} \bar{G}(\Vec{R},\Vec{k};i \omega_{n})  \right]_{-} \nonumber \\
& -\frac{i}{2} \left[ \Vec{\nabla}_{\Vec{R}} \bar{V}_{0}(\Vec{R},\Vec{k}_{\rm F})  \bar{\sigma}_{z}, \bar{\sigma}_{z} 
U_{\Vec{k}}^{M \dagger} \left[  \Vec{\nabla}_{\Vec{k}}  \check{G}(\Vec{R},\Vec{k},i \omega_{n}) \right] U_{\Vec{k}}^{M} 
\right]_{+} 
+i  \Vec{v}(\Vec{k}_{\rm F}) \cdot  \Vec{\nabla}_{\Vec{R}} \bar{G}(\Vec{R},\Vec{k};i \omega_{n}) = 0. \label{eq:el}
\end{align}
\end{widetext}
The $\xi_{\Vec{k}}$-integration erases the term with $\Vec{\nabla}_{\Vec{k}}$ in Eq.~(\ref{eq:el}), which 
 is same procedure in the single-band model. 
Thus, the resultant $2 M \times 2 M$ quasiclassical multi-band Eilenberger equation is

\section{Normalization condition}
\label{sec:normalization}
We consider the normalization condition for $\bar{g}$. 
When $\bar{g}$ satisfies Eq.~(\ref{eq:multi}), 
$\bar{g}' \equiv \alpha \bar{1} + \beta \bar{g}$ and $\bar{g} \bar{g}$ are also the solutions of  Eq.~(\ref{eq:multi}) as shown in Ref.~\cite{KopninText}. 
Thus, the solution of the equation has the form 
\begin{align}
\bar{g} \bar{g} &= \alpha \bar{1} + \beta \bar{g}.
\end{align}
The coefficients $\alpha$ and $\beta$ are determined in a homogeneous case. 
The $2 N \times 2 N$ matrix Green's function in a homogeneous system is written as\cite{Nagaisurface} 
\begin{align}
\check{G}(\Vec{k},i \omega_{n}) &= \check{U}(\Vec{k})
\left(\begin{array}{cc}\hat{A}_{+}(\Vec{k},i \omega_{n}) & \hat{B}(\Vec{k},i \omega_{n}) \\
\hat{B}^{\dagger}(\Vec{k},i \omega_{n}) & \hat{A}_{-}(\Vec{k},i \omega_{n})
\end{array}\right)
 \check{U}^{\dagger}(\Vec{k}), 
\end{align}
where 
%%%%%%%
\begin{align}
\left[ \hat{A}_{\pm}(\Vec{k},i \omega_{n}) \right]_{\alpha \beta} &= \delta_{\alpha \beta} \frac{i \omega_{n} \pm \epsilon_{\alpha}}{-|\Delta_{\alpha}|^{2} + (i \omega_{n})^{2} - \epsilon_{\alpha}^{2}}, \\
\left[ \hat{B}(\Vec{k},i \omega_{n})  \right]_{\alpha \beta} &=  \delta_{\alpha \beta} \frac{
\Delta_{\alpha}
}{
-|\Delta_{\alpha}|^{2} +(i \omega_{n})^{2} - \epsilon_{\alpha}^{2}.
}
\end{align}
%%%%%
Here, the $2 N \times 2 N$ matrix $\check{U}(\Vec{k})$ is the unitary matrix which diagonalizes $\check{H}^{N0}(\Vec{k})$. 
Assuming that intraband pairing are dominant, we neglect the off-diagonal (interband) elements of the order-parameter matrix. 

The $2 M \times 2 M$ Green's function in the projected space $\bar{G}(\Vec{k},i \omega_{n})$ is written as 
\begin{align}
 \bar{G}(\Vec{k},i \omega_{n}) &= \left(\begin{array}{cc}\tilde{A}_{+}^{M}(\Vec{k},i \omega_{n}) & \tilde{B}^{M}(\Vec{k},i \omega_{n}) \\
\tilde{B}^{M \dagger}(\Vec{k},i \omega_{n}) & \tilde{A}_{-}^{M}(\Vec{k},i \omega_{n})
\end{array}\right),
\end{align}
where 
\begin{align}
\left[ \tilde{A}_{\pm}^{M}(\Vec{k},i \omega_{n}) \right]_{\alpha \beta} &= \delta_{\alpha \beta} \frac{i \omega_{n} \pm \xi_{\Vec{k}}}{-|\Delta_{\alpha}|^{2} + (i \omega_{n})^{2} - \xi_{\Vec{k}}^{2}}, \\
\left[ \tilde{B}^{M}(\Vec{k},i \omega_{n}) \right]_{\alpha \beta} &= \delta_{\alpha \beta}  \frac{
\Delta_{\alpha}
}{
-|\Delta_{\alpha}|^{2} +(i \omega_{n})^{2} - \xi_{\Vec{k}}^{2}}.
\end{align}
By substituting the above Green's function into Eq.~(\ref{eq:quasi}), we obtain the multi-band quasiclassical Green's function 
in a homogeneous system expressed as
\begin{align}
\bar{g}(\Vec{k}_{\rm F},i \omega_{n}) &= 
\left(\begin{array}{cc}\tilde{a}_{+}^{M}(\Vec{k}_{\rm F},i \omega_{n}) & \tilde{b}^{M}(\Vec{k}_{\rm F},i \omega_{n}) \\
-\tilde{b}^{M \dagger}(\Vec{k}_{\rm F},i \omega_{n}) & -\tilde{a}_{-}^{M}(\Vec{k}_{\rm F},i \omega_{n})
\end{array}\right),
\end{align}
where 
\begin{align}
\left[ \tilde{a}_{\pm}^{M}(\Vec{k}_{\rm F},i \omega_{n}) \right]_{\alpha \beta} &= 
- \delta_{\alpha \beta} \pi  \frac{i \omega_{n}}{ \sqrt{\omega_{n}^{2} + |\Delta_{\alpha}|^{2}}}, \\
\left[ \tilde{b}^{M}(\Vec{k}_{\rm F},i \omega_{n}) \right]_{\alpha \beta} &= - \delta_{\alpha \beta} \pi  \frac{\Delta_{\alpha}}{ \sqrt{\omega_{n}^{2} + |\Delta_{\alpha}|^{2}}}.
\end{align}
Therefore, the normalization condition becomes Eq.(\ref{eq:normalization}) 
even in an inhomogeneous system. 
%%%%%%%

\end{document}